%% file: main.tex
\documentclass[preprint]{vgtc}               

\graphicspath{{figures/}{pictures/}{images/}{./}} 

\usepackage{times}                     

\usepackage{tabu}                      
\usepackage{booktabs}                  
\usepackage{lipsum}                    
\usepackage{mwe}                       

\usepackage{mathptmx}                  
\usepackage{graphicx}
\usepackage{pifont}
\usepackage{multirow}
\usepackage{amsmath}

\usepackage{enumitem}

\onlineid{1503}

\vgtccategory{Research}

\vgtcinsertpkg

\title{Exploring Avatar-Based Representations of Desktop Analytical Workflows for Asymmetric Collaborative Visual Analytics}

\author{Tiansu Chen\thanks{e-mail: tiansuchen@tamu.edu}\\ %
        \scriptsize Texas A\&M University %
\and Yalong Yang\thanks{e-mail: yalong.yang@gatech.edu}\\ %
     \scriptsize Georgia Tech %
\and Wai Tong\thanks{e-mail: wtong@tamu.edu}\\ %
     \scriptsize Texas A\&M University}

\teaser{
  \centering
  \includegraphics[width=0.79\linewidth]{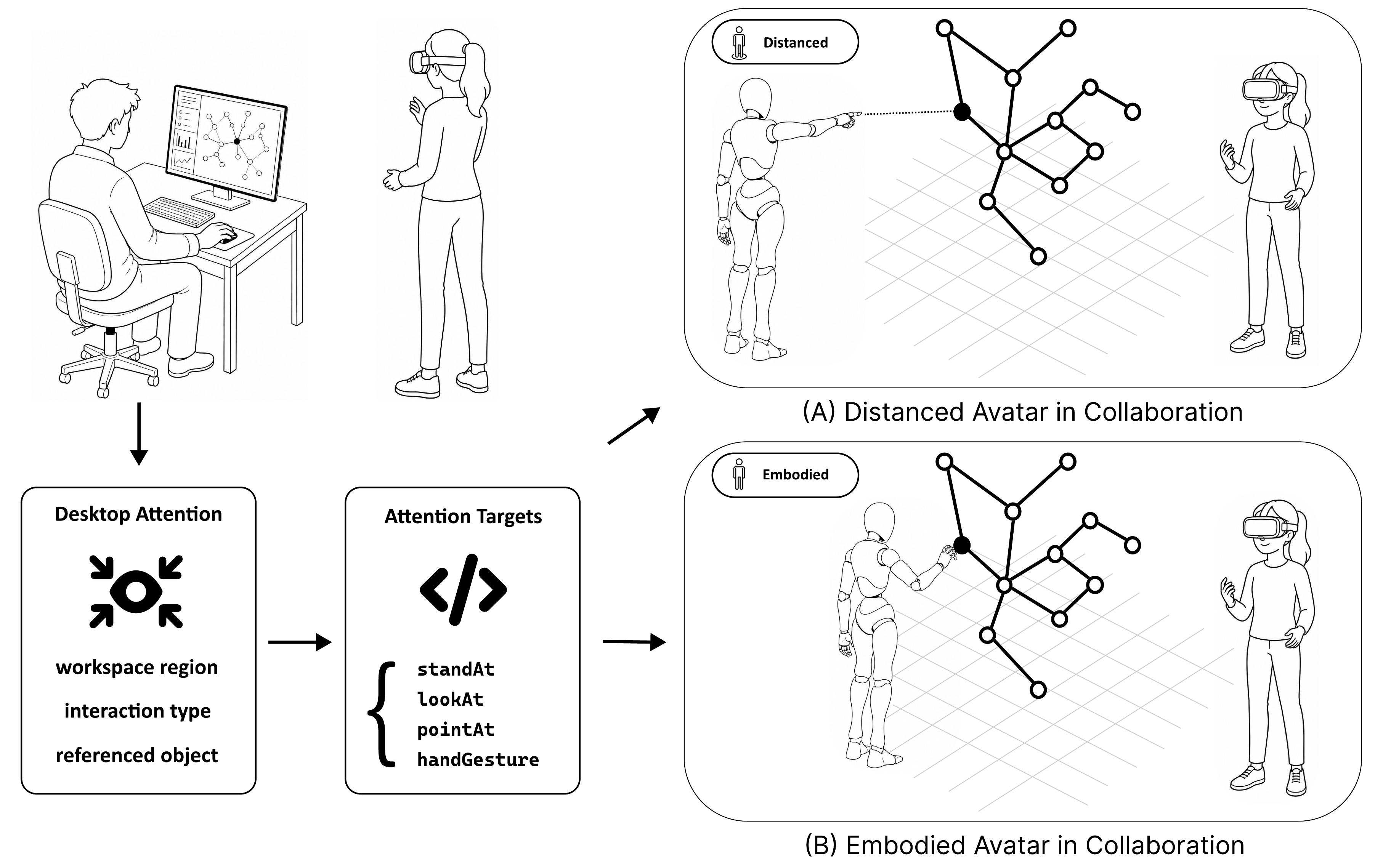}
  \caption{Overview of Desk2Avatar for representing desktop analytical activity in asymmetric desktop–VR collaboration. A desktop collaborator performs analytical actions on a node-link graph, which are represented in the VR workspace through avatar-based cues. We explore two avatar representations: (A) a Distanced Avatar that remains at a fixed position and indicates the collaborator’s focus through orientation and pointing, and (B) an Embodied Avatar that moves near target graph elements and uses close-range pointing to convey activity. The human figures and avatar depictions in this figure were generated using ChatGPT.}
  \label{fig:teaser}
}

\abstract{
Collaborative visual analytics increasingly occurs across asymmetric desktop–VR settings, with desktop analysis offering precision and efficiency and VR providing spatial and embodied affordances. However, maintaining workspace awareness remains challenging because desktop collaborators are often represented in VR only through indirect cues such as perspective sharing, shared visualization state, or lightweight cursor traces, which do not convey their ongoing analytical activity in a VR-native manner. To address this gap, we present Desk2Avatar, which explores avatar-based representations of desktop analytical workflows in VR. We introduce two representation strategies, Distanced and Embodied, inspired by remote pointing and direct manipulation. We conducted a within-subject study with 18 participants comparing these strategies against a depth-adaptive cursor baseline. Our findings show that avatar-based representations improved collaborator awareness and attentional guidance over cursor cues, while excessive embodiment introduced occlusion, distraction, and additional workload. Finally, we discuss design implications for future desktop-to-VR representations, focusing on balancing collaborator presence, attentional guidance, and workspace readability.
} 


\begin{document}


\input{sections/01_introduction}

\input{sections/02_related}

\input{sections/03_system}

\input{sections/04_evaluation}

\input{sections/05_result}

\input{sections/06_discussion}

\input{sections/07_limitation}

\input{sections/08_conclusion}

\bibliographystyle{abbrv-doi}

\bibliography{ref}
\end{document}

%% file: sections/01_introduction.tex
\firstsection{Introduction}

\maketitle

Collaborative visualization and visual analytics have become a key approach for data-intensive sensemaking, enabling distributed stakeholders to jointly explore complex datasets, construct shared interpretations, and coordinate analytical strategies through interactive visual representations \cite{isenberg2011collaborative, badam2017supporting}. Recent developments in immersive technologies have further expanded collaborative visual analytics by enabling users in immersive environments to participate in shared analytical activities alongside desktop collaborators, while engaging with data through spatial and embodied forms of exploration \cite{frohler2022survey}. This cross-platform setting has motivated growing interest in asymmetric collaborative visualization, in which participants engage in a shared analytical process across different devices and interaction contexts \cite{reski2022empirical, tong2023towards, enriquez2024evaluating, brehault2025systematic}.

However, this asymmetry introduces challenges for maintaining awareness of the collaborator's attention and activity while preserving a sense of their presence in the shared workspace. 
Prior work in collaborative visual analytics has underscored the importance of workspace awareness for supporting more tightly coupled collaboration \cite{gutwin2002descriptive, isenberg2011collaborative, lee2020shared, piumsomboon2019effects}, while studies of immersive collaboration have shown that co-presence and situated cues play an important role in shaping collaborative experience \cite{casanueva2000effects, schroeder2002social, oh2018systematic, bai2020user, yang2022towards}. 
These challenges are especially pronounced in desktop-VR collaboration because collaborators operate across fundamentally different 2D and 3D interaction contexts and workspaces. 
In this setting, representing desktop collaborators in VR is particularly difficult because desktop activity originates from screen-based interaction and lacks the spatial and embodied form native to immersive environments.

Existing work has therefore explored several approaches to representing desktop activity in VR, including perspective sharing, which presents the desktop user's viewpoint in a separate view \cite{saffo2023through, welsford2021spectator, seraji2022xvcollab}, and shared visualization state, which propagates selection and filtering states within the same view \cite{reski2022empirical, gottsacker2025examining}. 
While these techniques provide lightweight and device-agnostic support for coordination, they offer only indirect and discontinuous awareness of the desktop user, making it difficult for VR users to follow shifts in the collaborator's attention~\cite{gutwin2002descriptive, isenberg2011collaborative}.
Other work has provided more explicit representations of desktop activity through lightweight cursor-based cues \cite{tong2023towards, enriquez2024evaluating}, which convey pointer location and trajectory but provide limited detail about the underlying interactions. 
Across these approaches, desktop collaborators remain represented through abstract, non-embodied traces rather than as situated agents within the shared analytical space. 
As a result, such cues provide limited support for interpreting the analytical reasoning or evolving intent behind desktop actions, particularly in VR, where embodied and spatial signals play an important role in conveying attention, intention, and engagement \cite{sarasso2024shared}. Motivated by this gap, we asked (1) \textbf{how desktop analytical interactions can be translated into VR representations that convey a desktop collaborator's ongoing actions and analytical intent} and (2) \textbf{how different desktop-to-VR representations shape VR users' understanding of a desktop collaborator’s analytical process and their collaborative experience}.


To answer these research questions, we present \textbf{Desk2Avatar}, an asymmetric collaborative visualization system that translates desktop analytical interactions into avatar-based representations in VR. We introduce two representation modes, \textit{Distanced} and \textit{Embodied} (\autoref{fig:teaser}), motivated by two common interaction paradigms in VR: remote pointing and direct manipulation \cite{laviola20173d, mine1995virtual}. 
\textit{Distanced} keeps the avatar at a fixed position and conveys the desktop analyst’s current target through body orientation and pointing gestures toward the relevant visual elements. In contrast, \textit{Embodied} moves the avatar to the target element and represents the desktop interaction through direct touch-based gestures, such as reaching toward or touching the selected node in the immersive workspace. 
We evaluated how effectively avatar-based representations convey desktop interactions in VR through a user study with 18 participants against a depth-adaptive cursor baseline.
Our results indicate that avatar-based representations can improve the perceived presence of a desktop collaborator in VR, with the \textit{Distanced} avatar better supporting attention guidance and tracking the analytical process, while the \textit{Embodied} avatar showed the potential of spatially grounded body movement but was more affected by occlusion with visual elements and frequent movement across the workspace. 

Overall, our contributions are (1) a desktop-VR collaborative visualization prototype, Desk2Avatar, that conveys desktop analytical workflows in VR representations, (2) an empirical evaluation that examines the effects on performance, collaborative experience, and subjective preference, and
(3) design implications on avatar-based representations for desktop-VR collaborative visualization.


%% file: sections/02_related.tex
\section{Related Work}

\subsection{Asymmetric Collaborative Visual Analytics}

Asymmetric collaboration describes settings in which collaborators contribute through different devices, capabilities, perspectives, or roles \cite{brehault2025systematic}, and has been increasingly explored in visual analytics across heterogeneous workspace contexts. VisPorter~\cite{chung2014visporter} and PolyChrome~\cite{badam2014polychrome} explored collaborative visual analytics across multiple displays and devices, supporting information exchange and synchronized interactions across heterogeneous workspaces. In immersive analytics, this asymmetry extends across conventional and immersive displays, where collaborators may differ in interaction modality, spatial perspective, and level of immersion. Reski \textit{et al.} examined hybrid immersive and non-immersive analytics for synchronous data exploration, showing how collaborators establish shared context and make spatial or spatio-temporal references across interfaces~\cite{reski2020oh, reski2022empirical}. XVCollab~\cite{seraji2022xvcollab} explored collaborative analytics across the reality-virtuality continuum by allowing desktop and AR users to author, modify, and share visualizations from their respective environments.

Within asymmetric immersive analytics, several studies have specifically examined collaboration between desktop and VR devices. Saffo \textit{et al.} \cite{saffo2023through} highlighted the importance of awareness in desktop-VR collaborative analysis, while Tong \textit{et al.} \cite{tong2023towards} showed that PC-VR collaborative visualization can approach the effectiveness of symmetric settings despite distinct coordination trade-offs. Enriquez \textit{et al.} \cite{enriquez2024evaluating} also found that layout design influences performance and preference in PC-VR decision-making tasks, and Burova \textit{et al.} \cite{burova2022distributed} demonstrated the applicability of asymmetric immersive collaboration in distributed industrial workflows. Overall, prior work demonstrates the potential of asymmetric visual analytics while highlighting the challenges of communicating collaborators' activities across heterogeneous devices. While cross-device awareness has received increasing attention, how to effectively convey collaborators' actions, attention, and task context across different displays, perspectives, and interaction modalities remains an open design challenge.

\subsection{Cross-Device Workspace Awareness}


Workspace awareness is particularly important in cross-device collaboration, where partners work through different displays, viewpoints, and interaction modalities rather than a single shared interface \cite{gutwin2002descriptive}. 
In desktop-VR collaboration settings, supporting awareness of desktop users is more challenging, since their actions are mediated through mouse input, keyboard commands, and screen-based attention that are not naturally visible or readily interpretable in immersive environments.
In such, prior work has supported cross-device awareness through several recurring techniques. Some systems use perspective sharing to provide access to another collaborator's viewpoint, screen, or field of view \cite{piumsomboon2019effects, saffo2023through, seraji2022xvcollab}, while others reflect collaborators' actions directly within the shared workspace through shared visualization states, such as selections, highlights, annotations, or shared landmarks \cite{prouzeau2018awareness, reski2022empirical, rasmussen2022supporting, tong2023towards, saffo2023through}. Moreover, behavioral cues such as cursors, gaze, gestures, and pointing rays have also been used to convey collaborators' moment-to-moment attention and interaction position \cite{bai2020user, piumsomboon2019effects, yang2020effects, tong2023towards, woodworth2022redirecting, coppens2024supporting}.


However, while effective for communicating specific references or interaction positions, these non-embodied cues tend to present desktop activity as interface-level visual markers rather than as actions performed by a collaborator situated within the shared immersive workspace. This may weaken the perception of collaborator presence, which is important for making immersive collaboration feel socially grounded within a shared workspace \cite{casanueva2000effects, schroeder2002social, oh2018systematic, bai2020user, yang2022towards}.
Moving toward embodied representations, Woodworth \textit{et al.} \cite{woodworth2022redirecting} used a VR avatar to represent the desktop user in teaching-oriented scenarios by redirecting low-level inputs such as eye gaze, mouse pointing, and drawing into avatar animations. While their work focused on presenter-audience communication in remote VR meetings, our work studies asymmetric desktop-VR visual analytics, where avatar representations convey task-relevant interactions with data objects and the collaborator's evolving analytical process.

\subsection{Avatars in Immersive Analytics}

Avatar-based representation has been widely recognized as an effective form of user representation in VR, because it makes users perceptible to one another through bodily, spatial, and socially interpretable cues \cite{oh2018systematic}. Prior work has shown that avatar embodiment, nonverbal expressiveness, and spatial co-presence shape how users perceive and interpret others in immersive environments, particularly when inferring collaborators' focus, intentions, and ongoing activity from limited interaction cues \cite{kyrlitsias2022social, wei2022communication}. 
In collaborative immersive analytics, avatars become more directly tied to analytical activity and workspace awareness. Prior systems have used shared immersive spaces and avatar-based cues to support collaborative data exploration and coordination around visualizations \cite{dwyer2018immersive, donalek2014immersive}. More recently, Lee \textit{et al.} \cite{lee2020shared} showed that avatars and pointers supported deixis and coordination when collaborators jointly created, arranged, and interpreted visualizations, while collaborative graph systems have similarly used avatar representations such as floating heads and hands to support multi-user exploration and awareness \cite{heidrich2021towards}. These examples indicate that avatars in immersive analytics can support analytical awareness by conveying collaborators' spatial positions, attention, and interactions with shared visualizations.


In contrast, asymmetric collaborative immersive analytics introduces a distinct representational challenge because collaborators may differ in interface, input modality, and level of immersion. Zagermann \textit{et al.} identified transitional user representations as a key challenge in hybrid immersive analytics, ranging from cursors in non-immersive views to realistic avatars in immersive environments \cite{zagermann2023challenges}. In desktop-VR settings, Hart \textit{et al.} compared in-scene 3D avatars with screen-attached 2D avatars for communication and social presence \cite{hart2021manipulating}, while Woodworth \textit{et al.} mapped desktop inputs such as mouse pointing, gaze, and gestures onto cross-reality avatars for presentation scenarios \cite{woodworth2022redirecting}. However, these studies focused primarily on interpersonal communication and presentation rather than representing the spatial and sequential structure of desktop analytical activity. Our work addresses this gap by examining how desktop analytical interactions can be translated into avatar-based VR representations and how different strategies affect users' understanding of the desktop collaborator's analytical process.

%% file: sections/03_system.tex
\section{Desk2Avatar}
 
Adapting from prior studies on collaborative visualization~\cite{balakrishnan2008visualizations, mahyar2014supporting, tong2023towards}, we chose a node-link graph-based problem-solving scenario because it supports externalizing analytical reasoning through entity relationships~\cite{gou2012socialnetsense} and provides a meaningful setting for spatial graph exploration in VR~\cite{belcher2003using, ware1994viewing, ware1996evaluating}. 

\subsection{Desktop-to-VR Representation Pipeline}


Desk2Avatar uses a three-stage pipeline (\autoref{fig:teaser}) to translate desktop analytical activity into VR representations. It first resolves attention targets from document inspection and graph interaction, remaps these targets into VR space, and then represents them through avatar behaviors. The depth-adaptive cursor baseline is integrated into the same pipeline to ensure a unified translation process across conditions.


\vspace{0.3em}\noindent\textbf{Resolving Desktop Attention.} The first stage resolves desktop activity into an attention target, an intermediate representation of the collaborator's current analytical focus. Inspired by the structured interaction model proposed by Zhao \textit{et al.}~\cite{zhao2025libra}, we 
define the attention target as a structured description of the activity, capturing the workspace region in which the activity occurs, the type of interaction being performed, and the referenced object or position involved in the interaction.
Based on this structure, 
Desk2Avatar resolves attention differently for document-space and graph-space activity. In the source document space, the system identifies whether the desktop collaborator is browsing the document collection or focusing on a specific document. When a document is selected, it becomes the referenced target, while any text selection further refines the activity as focused evidence inspection. In the graph space, the system resolves attention by combining the interaction type with the involved graph element or position. For example, selecting a node combines a selection action with the selected node, while adding a link combines a construction action with the relationship being created. For graph navigation, 
we convert the desktop cursor into a device-agnostic graph position. Adapted from cross-device cursor approaches~\cite{tong2023towards,zhou2022depth}, the system identifies nearby reference nodes, computes their relative influence, and uses them to reconstruct the corresponding position in graph space.


\vspace{0.3em}\noindent\textbf{Remapping Attention Targets.} After resolving desktop attention, the second stage remaps the structured attention target into VR space. 
While the attention target captures what the desktop collaborator is attending to, it does not specify how that focus should be expressed through an avatar in an immersive workspace. Prior work on embodied agents and avatar-mediated communication has shown that gaze, posture, body orientation, gesture, and spatial positioning are central channels for conveying attention, reference, and social presence in virtual environments~\cite{cassell2001embodied, oh2018systematic}. Building on avatar retargeting research that separately models avatar placement, head orientation, and deictic gestures across virtual spaces~\cite{yoon2021full, yang2024visual}, we organize these channels into four independently controllable parameters: \texttt{standAt}, \texttt{lookAt}, \texttt{pointAt}, and \texttt{handGesture}.
\texttt{standAt} and \texttt{lookAt} are derived from the workspace region and the referenced target to specify where the avatar may be situated and where its attention is directed, respectively. \texttt{pointAt} is assigned to selection interactions and uses the referenced object as the pointing target, while \texttt{handGesture} is derived from the interaction type to express short interaction events such as graph construction or revision.

\begin{table}[t]
\centering
\caption{Characteristics of three VR representations.}
\label{tab:representation_characteristics}
\renewcommand{\arraystretch}{1.15}
\resizebox{\columnwidth}{!}{
\begin{tabular}{l|ccc}
\hline\hline
\textbf{Dimension} & \textbf{Distanced Avatar} & \textbf{Embodied Avatar} & \textbf{Depth Cursor} \\
\hline\hline

\textbf{Design paradigm} & Remote pointing & Direct manipulation & Spatial marker \\
\hline\hline

\textbf{Attention cues} & & & \\
\quad Avatar relocation & \ding{55} & \ding{51} & \ding{55} \\
\quad Body orientation & \ding{51} & \ding{51} & \ding{55} \\
\quad Target pointing & \ding{51}, distant & \ding{51}, near-target & \ding{55} \\
\quad Indication ray & \ding{51} & \ding{55} & \ding{55} \\
\hline\hline

\textbf{Parameterization} & & & \\
\quad \texttt{standAt} & \ding{55} & \ding{51} & \ding{55} \\
\quad \texttt{lookAt} & \ding{51} & \ding{51} & \ding{51} \\
\quad \texttt{pointAt} & \ding{51} & \ding{51} & \ding{55} \\
\quad \texttt{handGesture} & \ding{51} & \ding{51} & \ding{55} \\
\hline\hline

\end{tabular}
}
\end{table}


\begin{figure*}[t]
    \centering
    \includegraphics[width=0.9\textwidth]{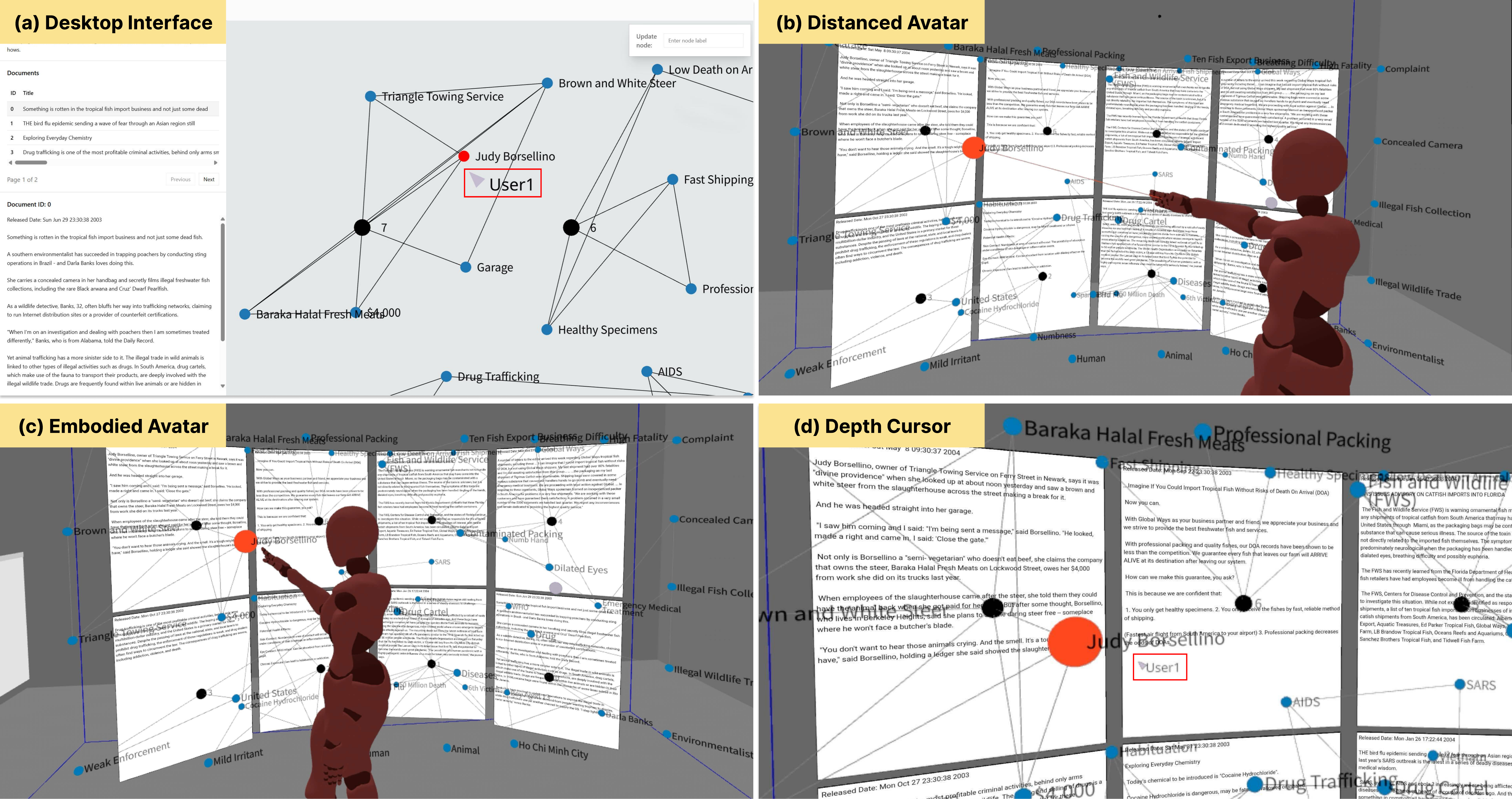}
    \caption{Different VR representations of a desktop node selection. (a) The desktop collaborator selects a node in the node-link graph space. The red box indicates the current cursor position and area of interest. (b) The Distanced avatar points toward the corresponding VR node from a fixed position using a controller-like ray. (c) The Embodied avatar moves toward the node and indicates the selection with a near-touch pose. (d) The depth-adaptive cursor places a cursor marker, shown in the red box, at the remapped target position. Across all conditions, the selected node is enlarged and highlighted in red.}
    \label{fig:representation}
\end{figure*}

\vspace{0.3em}\noindent\textbf{Representations in VR.} In the final stage, the system turns the remapped parameters into different VR representations. Inspired by two common VR interaction paradigms, remote pointing and direct manipulation~\cite{laviola20173d, mine1995virtual}, we designed two corresponding avatar-based representation modes, \textit{Distanced} and \textit{Embodied}. 
We also integrated a depth-adaptive cursor baseline, providing a non-avatar spatial cue derived from the same remapped parameters. \autoref{fig:representation} illustrates how the same desktop node selection is represented across the three conditions, while Table~\ref{tab:representation_characteristics} summarizes the main characteristics.

\begin{itemize}[leftmargin=*, nosep]
\item\textbf{Distanced Avatar.} The Distanced avatar represents the desktop collaborator from a fixed position in VR and therefore does not use \texttt{standAt} to update its location. Instead, it expresses the remapped target from a distance through orientation, pointing, and brief action cues. \texttt{lookAt} orients the avatar toward the relevant target, while \texttt{pointAt} indicates selected objects with one or both hands. During node selection (\autoref{fig:representation}(b)), for example, the avatar points toward the corresponding node from its fixed position, with an indication ray connecting the hand and target to make this remote reference more explicit. For graph editing activities, \texttt{handGesture} triggers a short hand or arm animation.

\item\textbf{Embodied Avatar.} The embodied avatar represents the desktop collaborator through relocation and near-target actions in VR. Rather than remaining fixed, it uses the \texttt{standAt} parameter to move toward the remapped target through a walking animation. Once near the target, \texttt{lookAt} orients the avatar toward the relevant document, graph region, or interaction object, while \texttt{pointAt} guides reaching toward selected objects. In the node selection example, the avatar approaches the corresponding node and indicates it with a near-touch pose as shown in \autoref{fig:representation}(c). For graph editing, \texttt{handGesture} triggers short near-target hand or arm animations associated with the interaction type, allowing the action to appear localized around the target.

\item\textbf{Depth-adaptive Cursor (baseline).} We implemented a depth-adaptive cursor baseline following Tong \textit{et al.}~\cite{tong2023towards}, representing desktop activity as a lightweight, non-embodied spatial cue in VR. The cursor is positioned according to the remapped \texttt{lookAt} target and updates as the target changes. Interacted graph objects are highlighted without additional avatar-based cues. \autoref{fig:representation}(d) shows the cursor representation for node selection.
\end{itemize}

\subsection{System Interface}

\begin{figure*}[htbp]
    \centering
    \includegraphics[width=\textwidth]{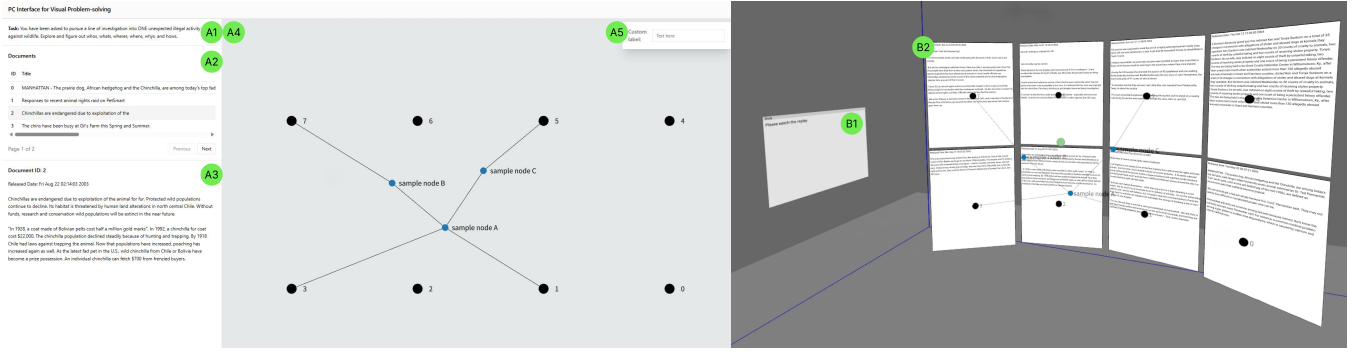}
    \caption{System interface. The desktop interface (left) mainly consists of five parts: the task description (A1), the document list (A2), the selected document (A3), the node-link graph (A4), and the label input box (A5). The VR interface (right) mainly consists of two parts: the information panel (B1) and the integrated document and node-link graph space (B2).}
    \label{fig:interface}
\end{figure*}

We adapt the interface design of Tong \textit{et al.}~\cite{tong2023towards}.
\autoref{fig:interface} shows the interface on both the desktop and VR sides. On the desktop side, users read documents, extract relevant information, and organize their findings into a node-link graph, while the task description (A1) provides a task overview, and the document list (A2) and selected document view (A3) support navigation and inspection of the source documents. Based on the entities, events, and relationships identified from the documents, users can create graph elements in the node-link graph space (A4), where each source document is represented by a black document node. New nodes are created by highlighting source text and clicking in the graph space, with the selected text used as the node label and optionally edited through the label input box (A5), and each newly created node is automatically connected to the document node of the active document.

On the right side of \autoref{fig:interface}, the VR interface presents the documents and graph elements in a unified spatial workspace for immersive sensemaking. The information panel (B1) provides task description and study instructions, while the integrated document and node-link graph space (B2) arranges source documents around the user in a semi-circular layout, allowing multiple documents to remain visible simultaneously. Each document is paired with a corresponding document node in front of it, spatially anchoring the graph structure to the document layout. Although the VR interface supports basic controller-based interactions graph editing, it primarily serves as the receiving environment for desktop-to-VR representations in this work, with VR users are only required to select graph nodes during the user study.

\subsection{Implementation}

Desk2Avatar was implemented as a web-based desktop-VR collaborative visualization application using React.js, D3.js, Three.js, and WebXR, with a Node.js and WebSocket-based client-server architecture for synchronizing document, graph, interaction, and layout states across clients. The system computes the 2D and 3D node-link graph layouts using d3-force-3d\footnote{\url{https://github.com/vasturiano/d3-force-3d}}. Avatar models and base animations were obtained from Adobe Mixamo\footnote{\url{https://www.mixamo.com}}, with skeletal animation and inverse-kinematics-based arm control used to represent desktop activities in VR. We will release the source code upon acceptance.

%% file: sections/04_evaluation.tex
\section{User Study}
We conducted a within-subject user study to investigate how different representations of desktop collaborators affect VR users' understanding of collaborator's actions and analytical intent, task performance and experience. We also varied task complexity to examine whether these effects changed with task demands. This study was approved by the Institutional Review Board, and all participants provided informed consent.

\subsection{Study Design}

As we focus on desktop-to-VR representation, we constrain the collaborative setting to uninterruptible, unidirectional interaction to isolate how different representations affect VR users' understanding of collaborator actions, task performance, and overall experience. 
To ensure consistency across conditions, we replayed a fixed set of pre-recorded desktop actions in VR, keeping collaborator behavior and conveyed content constant across trials. This design minimizes variability from live collaboration while keeping collaborator actions and conveyed content constant across trials, enabling controlled evaluation of representation effects.

\vspace{0.2em}\noindent\textbf{Task.} Based on the graph task taxonomy \cite{lee2006task}, we selected browsing tasks because they require users to follow sequential collaborator actions and maintain awareness of the exploration process, making them suitable for evaluating representation effects.
Participants viewed a short pre-recorded replay in which the experimenter, acting as the desktop collaborator, browsed a node-link graph while narrating a brief analytical thread. The desktop activity consisted of a node-selection sequence, with each node representing an entity, event, location, activity, or concept. \autoref{fig:representation} illustrates one replay step from the VR perspective across the three representation conditions. The node selections were synchronized with audio narration explaining their relationships and contributions to the analytical thread. After each replay, participants reproduced the selected-node sequence and reported their understanding of the analysis.

\vspace{0.2em}\noindent\textbf{Dataset.} We used the Blue Iguanodon dataset from VAST 2007 \cite{grinstein2007vast}, a document-based visual analytics dataset. It contains reports describing interconnected entities, events, and suspicious activities, supporting graph-based analytical narratives. We extracted three threads involving drug trafficking, wildlife smuggling, and bioterrorism. Each thread included a primary crime-related storyline with multiple entities, relationships, and events, along with shorter secondary storylines from the same information context. We used the secondary storylines for easy tasks and the primary storylines for hard tasks. For each thread, we selected six documents supporting both storyline types and added two distractor documents to introduce irrelevant entities and relationships, better approximating a realistic analytical environment with noisy information.

\vspace{0.2em}\noindent\textbf{Graph.} For the eight documents in each thread, we constructed a node-link graph to represent the corresponding information space. We identified key entities, events, and concepts from the primary and secondary storylines as nodes, and created links from explicit relationships described in the documents. We also connected nodes when relationships emerged across documents to preserve narrative structure and support cross-document reasoning. To ensure comparable materials across threads, we controlled graph size while preserving the key event and relationship structures. The resulting graphs were similar in scale, containing 56 nodes and 98 links, 58 nodes and 101 links, and 61 nodes and 106 links, respectively.

\begin{table}[t]
    \centering
    \caption{Task complexity parameters for the graph browsing tasks.}
    \label{tab:task_complexity}
    \begin{tabular}{llccc}
        \toprule
        Thread & Level & Nodes & Revisits & Jumps \\
        \midrule
        \multirow{2}{*}[-0.2em]{Thread 1} & Easy & 10 & 0 & 0 \\
        \cmidrule(){2-5}
                                  & Hard & 20 & 2 & 2 \\
        \midrule
        \multirow{2}{*}[-0.2em]{Thread 2} & Easy & 11 & 0 & 0 \\
        \cmidrule(){2-5}
                                  & Hard & 19 & 1 & 2 \\
        \midrule
        \multirow{2}{*}[-0.2em]{Thread 3} & Easy & 9  & 0 & 0 \\
        \cmidrule(){2-5}
                                  & Hard & 19 & 2 & 1 \\
        \bottomrule
    \end{tabular}
\end{table}

\vspace{0.2em}\noindent\textbf{Complexity.} Since prior work has shown that visual encoding effects can vary with task difficulty \cite{guo2023effects}, we introduced two complexity levels to examine whether collaborator representations were similarly affected as the graph browsing became more demanding.
We varied complexity using two structural features, \textbf{revisit} and \textbf{jump}, building on graph task taxonomies \cite{lee2006task}.
A revisit returns to a previously selected node, while a jump moves to a node not directly connected to the current node, reflecting transitions across documents or subtopics.
Easy tasks followed short, connected paths with linear action sequences, and hard tasks were longer and included revisits and jumps, requiring participants to remember prior nodes and interpret non-linear transitions in the collaborator's narrative.
Detailed task complexity parameters are summarized in Table~\ref{tab:task_complexity}.

\subsection{Participants}

We recruited 18 participants (7 females, 11 males) through recruitment emails. All were at least 18 years old (7 aged 18–24, 11 aged 25–34), including 4 bachelor's, 9 master's, and 5 doctoral students. Participants were screened for safe VR participation and reported normal or corrected-to-normal vision and sufficient English proficiency. VR experience ranged from more than two years (5), one to two years (2), and one month to one year (7), to no prior experience (3), with 1 participant preferring not to report. For familiarity with interactive data visualizations, participants reported being slightly familiar (6), moderately familiar (4), very familiar (4), or extremely familiar (2), while 2 preferred not to report.

\subsection{Procedure}
Each session lasted approximately 90 minutes, and a \$20 Amazon eGift card was given to each participant as compensation.

\vspace{0.2em}

\noindent\textbf{Introduction ($\sim$5 minutes).} Participants were welcomed and given an overview of the study purpose, procedure, and experimental setup. They were then asked to review and sign an informed consent form before the session began. After providing consent, participants completed a brief demographics questionnaire.

\vspace{0.2em}

\noindent \textbf{Training ($\sim$10 minutes).} Participants completed a brief training session to become familiar with the study setup, collaborator representations, and the basic interaction of the replay-based graph browsing task. The experimenter guided participants through the replay and response process, allowing them to understand the VR viewing experience, the representation cues, and the general form of node-link graph interaction before beginning the main study.

\vspace{0.2em}

\noindent \textbf{Main Study ($\sim$65 minutes).} Participants completed two task rounds at different complexity levels, each containing three conditions: two avatar-based representations and one non-embodied baseline. The order of conditions was counterbalanced using a balanced Latin square design \cite{bradley1958complete}, while the dataset sequence was controlled across rounds. 
For each condition, the experimenter started the replay, after which all selections were cleared. Participants then reproduced the desktop collaborator's node-selection sequence and answered questions assessing their understanding of the analysis, perceived workload using NASA-TLX \cite{hart2006nasa}, and co-presence \cite{biocca2001networked}.
After all three conditions in a round, participants completed a questionnaire ranking the conditions on several criteria.

\vspace{0.2em}

\noindent \textbf{Debriefing ($\sim$10 minutes).} After the main study, participants completed a short semi-structured interview about their experience. The interview focused on overall impressions, perceived engagement, and how different representations supported their understanding of the desktop collaborator's actions and analytical content.

\subsection{Measures}

To examine how different desktop-to-VR representations shape VR users' understanding of a desktop collaborator's analytical process and their collaborative experience,
we organized our measures around two aspects: \textbf{performance} and \textbf{awareness}. Specifically, performance captures how effectively users follow and understand the graph-based analytical process, while awareness reflects their perception of the collaborator and collaborative experience.

For \textbf{performance}, we focused on analytical recall and perceived workload. Analytical recall included accuracy in reproducing the collaborator's node-selection sequence and understanding of the conveyed analysis. Sequence recall was computed by comparing each participant's reconstructed sequence with the ground-truth replay using a normalized Levenshtein Distance \cite{levenshtein1966binary}:

\begin{equation}
    \text{Recall Score} = 1 - \frac{\text{Levenshtein Distance}}{\max\left(|S_{\text{groundtruth}}|, |S_{\text{participant}}|\right)}
\end{equation}

where $S_{\text{groundtruth}}$ and $S_{\text{participant}}$ denote the original and reconstructed node sequences, respectively. This measure captures omissions, additions, ordering errors, and repeated visits. 
To assess analytical understanding beyond sequence recall, we scored participants' responses using rubrics based on key entities, relationships, events, and causal connections conveyed in each replay. Perceived workload was measured using NASA-TLX \cite{hart2006nasa} to capture the effort required to follow and interpret the collaborator's actions under each representation.
For \textbf{awareness}, we used the Networked Minds Measure of Social Presence \cite{biocca2001networked} to assess participants' awareness of and perceived connection with the desktop collaborator during the task. Moreover, we collected comparative preferences and post-study interview feedback to complement these measures and provide subjective accounts of participants' experiences across representation conditions.

%% file: sections/05_result.tex
\section{Results}

\begin{figure}[t]
    \centering
    \includegraphics[width=\columnwidth]{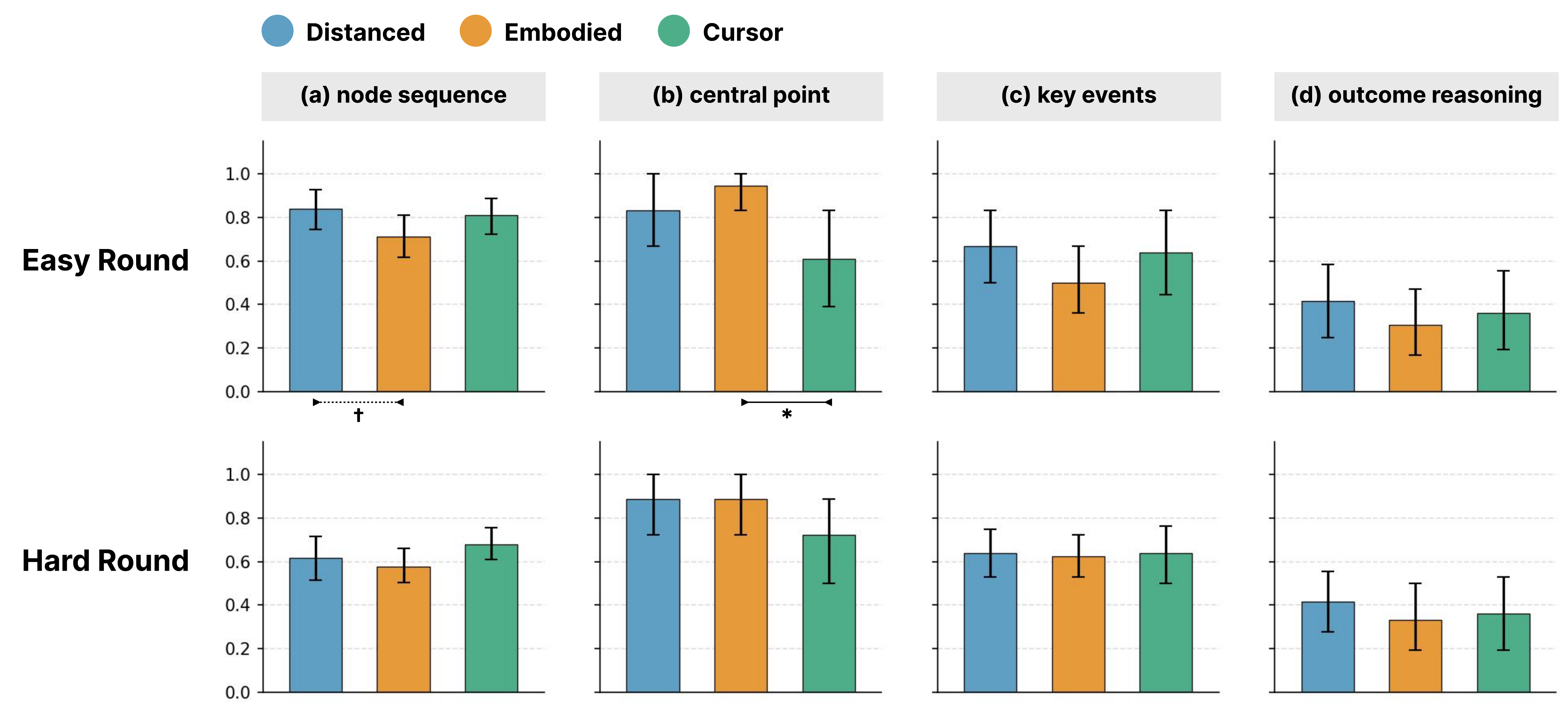}
    \caption{Performance comparison across two task rounds. (a) Node sequence reconstruction accuracy. (b)–(d) Scores for analysis comprehension Q\&A about the central point, key events, and outcome reasoning. 
    Bars show means with 95\% confidence intervals. Significance annotations: $*$ for $0.01 \leq p < 0.05$, $\dagger$ for $0.05 \leq p < 0.10$.}
    \label{fig:performance}
\end{figure}

We analyzed all measures using non-parametric statistical tests. For each task round, we conducted a Friedman test to compare the three representation conditions separately for easy and hard tasks. Significant effects were followed by pairwise Wilcoxon signed-rank tests with Holm-Bonferroni correction~\cite{holm1979simple} to control Type I error from multiple comparisons. We selected this step-down correction because it is less conservative than standard Bonferroni correction, making it better suited for our exploratory analysis~\cite{aickin1996adjusting}. To examine effects of task complexity, we also used Wilcoxon signed-rank tests within each representation condition to compare easy and hard rounds. For significant pairwise comparisons, we reported the absolute matched-pairs rank-biserial correlation ($|r_{rb}|$) as effect size \cite{kerby2014simple}. We used $\alpha = 0.05$ for statistical significance and reported $0.05 < p < 0.10$ as suggestive trends to avoid strictly dichotomous interpretation of p-values \cite{wasserstein2016asa, dragicevic2016fair, schumm2013determining}.

\subsection{Performance}


\vspace{0.2em}\noindent\textbf{Analytical Recall and Comprehension.} The task performance comparison is shown in \autoref{fig:performance}.
In the easy round, participants scored significantly higher on the (b) central point summarization with the Embodied avatar than with the Cursor ($p = 0.0429$, $|r_{rb}| = 1.000$), while no significant differences were found in the hard round. In addition, the figure also shows a tendency for (a) node sequence reconstruction accuracy to be generally lower in the hard round than in the easy round across conditions, while Q\&A scores generally decreased from (b) central point to (c) key events and (d) outcome reasoning within each round.


\begin{figure}[t]
    \centering
    \includegraphics[width=\columnwidth]{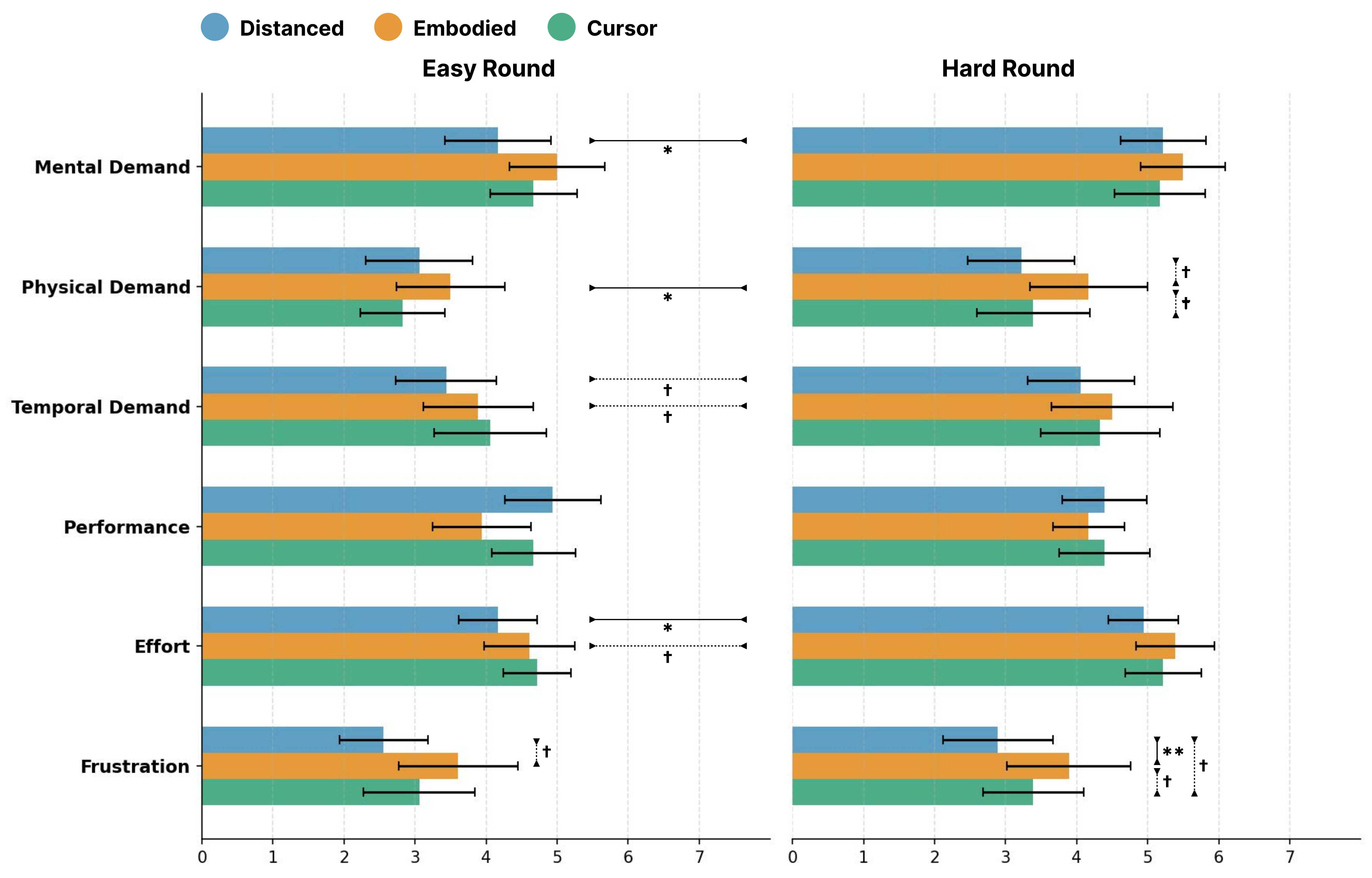}
    \caption{NASA-TLX result across two task rounds. 
    Bars show means with 95\% confidence intervals. Significance annotations: $**$ for $0.001 \leq p < 0.01$, $*$ for $0.01 \leq p < 0.05$, $\dagger$ for $0.05 \leq p < 0.10$.}
    \label{fig:tlx}
\end{figure}

\vspace{0.2em}\noindent\textbf{Workload.} For the NASA-TLX results shown in \autoref{fig:tlx}, we found no significant differences among the three representation conditions in the easy task round. In the hard task round, participants reported significantly lower frustration in the Distanced avatar condition than in the Embodied avatar condition ($p=0.008$, $|r_{rb}| = 1.000$).
Comparisons across task rounds showed that, with the Distanced avatar representation, participants reported significantly higher mental demand ($p=0.0253$, $|r_{rb}| = 0.633$) and effort ($p=0.0129$, $|r_{rb}| = 0.855$) in the hard task round than in the easy task round. With the Embodied avatar representation, participants reported significantly higher physical demand in the hard task round than in the easy task round ($p=0.0348$, $|r_{rb}| = 0.778$).

\subsection{Awareness}


\begin{figure*}[t]
    \centering
    \includegraphics[width=0.9\textwidth]{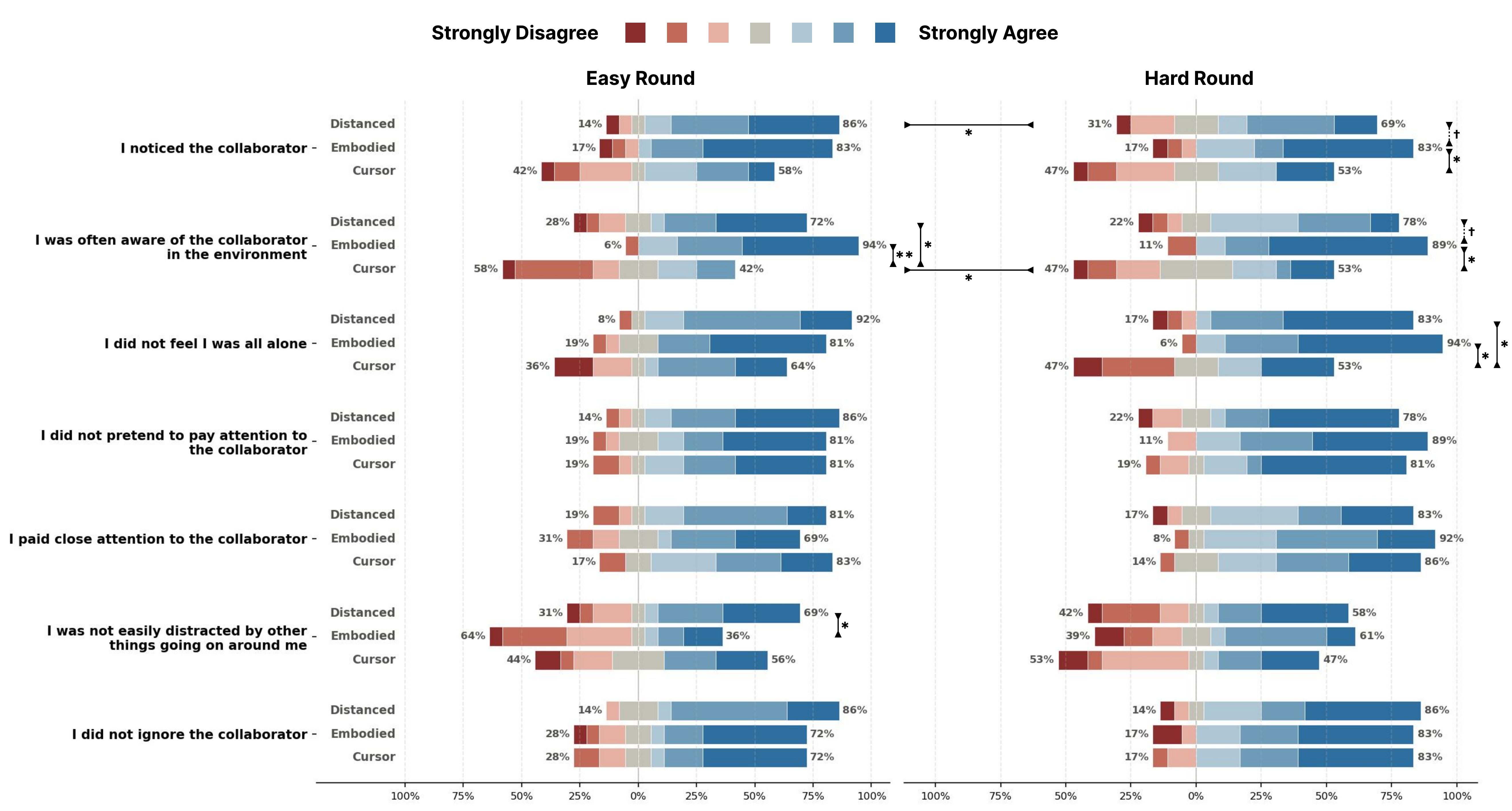}
    \caption{Co-presence ratings across two task rounds
    Scores for selected items were reversed so that higher values consistently indicate more positive responses. 
    Significance annotations: $**$ for $0.001 \leq p < 0.01$, $*$ for $0.01 \leq p < 0.05$, $\dagger$ for $0.05 \leq p < 0.10$.}
    \label{fig:copresence}
\end{figure*}

The co-presence ratings are shown in Figure~\ref{fig:copresence}. In the easy round, collaborator awareness was significantly higher with both the Embodied ($p=0.0025$, $|r_{rb}|=0.975$) and Distanced avatars ($p=0.0477$, $|r_{rb}|=0.640$) than with the Cursor. Participants also reported significantly less distraction with Distanced avatar than the Embodied avatar ($p=0.0213$, $|r_{rb}| = 0.775$). In the hard round, the Embodied avatar scored significantly higher than the Cursor for noticing the collaborator ($p=0.0435$, $|r_{rb}| = 0.733$), collaborator awareness ($p=0.0108$, $|r_{rb}| = 0.876$), and not feeling alone ($p=0.0119$, $|r_{rb}| = 0.936$). The Distanced avatar also scored higher than the Cursor for not feeling alone ($p=0.0191$, $|r_{rb}|=0.879$).

Across rounds, scores for noticing the collaborator were significantly lower with the Distanced avatar in the hard round than the easy round ($p=0.0266$, $|r_{rb}| = 0.670$). In contrast, collaborator awareness with the Cursor was significantly higher in the hard round than the easy round ($p=0.0480$, $|r_{rb}| = 0.615$). No significant across-round differences were found for the Embodied avatar.

\subsection{Preference}

\begin{figure*}[t]
    \centering
    \includegraphics[width=0.8\textwidth]{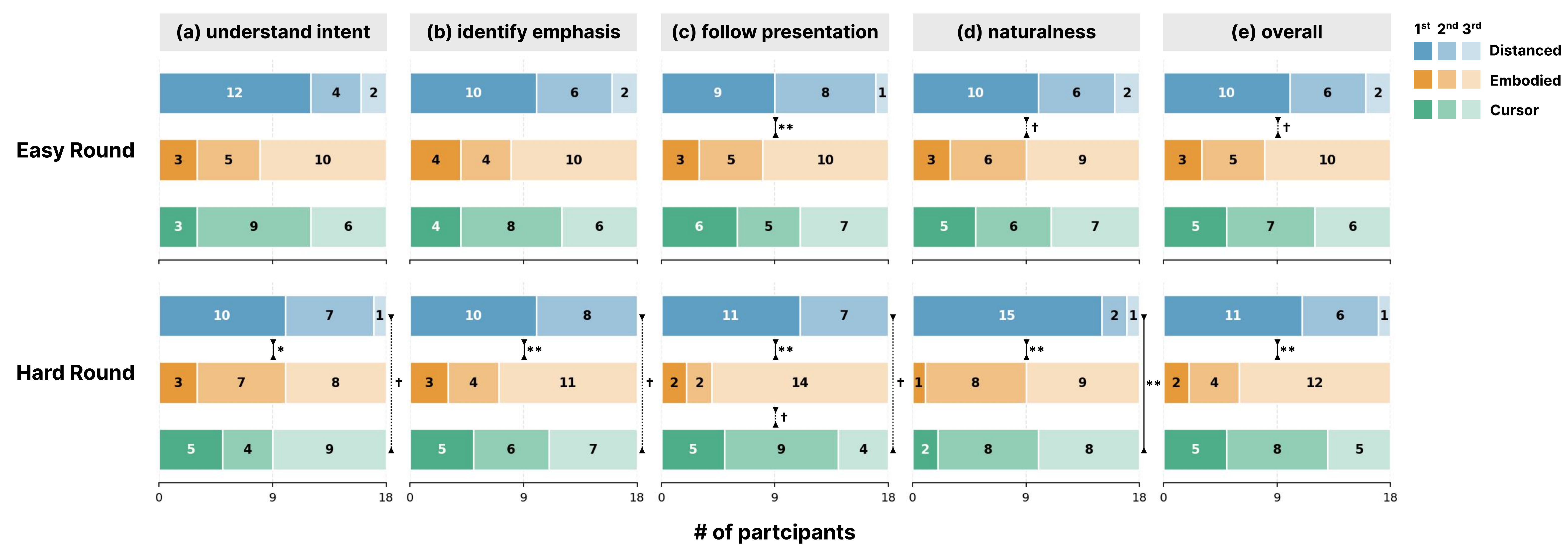}
    \caption{Preference ranking distributions across two task rounds. Participants ranked the three conditions based on (a) understanding intent, (b) identifying emphasis, (c) following the presentation, (d) naturalness, and (e) overall preference. 
    Significance annotations: $**$ for $0.001 \leq p < 0.01$, $*$ for $0.01 \leq p < 0.05$, $\dagger$ for $0.05 \leq p < 0.10$.}
    \label{fig:ranking}
\end{figure*}

Figure~\ref{fig:ranking} shows participants’ preference rankings across five criteria in the two task rounds. In the easy round, the Distanced avatar ranked significantly higher than the Embodied avatar for following the presentation ($p=0.0337$, $|r_{rb}| = 0.649$). In the hard round, Distanced ranked significantly higher than Embodied across all five criteria: understanding intent ($p=0.0312$, $|r_{rb}| = 0.649$), identifying emphasis ($p=0.0084$, $|r_{rb}| = 0.772$), following the presentation ($p=0.0021$, $|r_{rb}| = 0.883$), naturalness ($p=0.0025$, $|r_{rb}| = 0.871$), and overall preference ($p=0.0058$, $|r_{rb}| = 0.807$). Distanced avatar also ranked higher than the Cursor for naturalness ($p=0.0066$, $|r_{rb}| = 0.766$).

\subsection{Qualitative Feedback}

We used the affinity-diagramming approach \cite{hartson2012ux} to analyze participants' interview transcripts. 

\vspace{0.2em}\noindent\textbf{Avatar Representations.} Participants reported two shared benefits of the avatar representations. First, four participants (P2, P10, P13, and P14) reported that avatars strengthened the perceived co-presence with the desktop collaborator. P2 described the avatar as \textit{``a companion''} that helped them feel \textit{``as if I'm not working alone.''} while P13 commented it felt like \textit{``there is an actual presenter being there with me instead of just a background sound.''}.
Second, nine participants (P1-3, P6, P8, P10, P12, P15, and P16) found that body movement and pointing made shifts in collaborator attention easier to perceive. P1 noted that the avatar could \textit{``point or just run over to help you find where a point jumped to.''} while P16 found its movement \textit{``more conspicuous than the cursor.''} 

\vspace{0.2em}

\noindent \textbf{Distanced Avatar.} Participants generally found the Distanced avatar helpful for guiding attention while remaining outside the main graph space. Ten participants (P2, P3, P5-8, P10, P13, P15, and P16) commented that its arm and pointing ray made targets easier to follow, especially across node transitions in hard tasks. P2 described it as \textit{``really helpful to kind of like direct my view whenever it was jumping from node to node,''} while P13 explained that \textit{``you can trace the line, that goes from the avatar to the nodes, but that's not for the cursor one.''}
Six participants (P3, P6, P7, P12, P13, and P18) also described its side placement as resembling a presenter guiding attention from outside the graph. P6 stated that it felt like \textit{someone next to me just pointing where to look,''} and P18 commented that \textit{having a person there made it easier to understand what they are trying to say.''}
However, participants also noted two limitations. Three participants (P3, P9, and P17) found that its bodily presence and arm movements could still distract from the graph. P9 mentioned that \textit{``I could still see the avatar moving ... because I also get distracted with a lot of [upper-body and arm] movement.''} Two participants (P6 and P12) also found the pointing cue less clear in dense graph areas. P12 explained that \textit{``as the tasks compiled a lot for the Distanced one, you couldn't see as much since there was a point behind another point.''}

\vspace{0.2em}

\noindent\textbf{Embodied Avatar.} Participants reported that the Embodied avatar made the collaborator activity more visible by moving toward pointing directly at relevant graph elements. 
Several participants (P1, P6-8, P12, P13, and P15) described this useful for locating the target and following the analytical flow. P8 noted that, when the flow became \textit{``longer and complicated,''} the Embodied avatar made it \textit{``easier to understand the overall plot that was going on.''} P12 also found it useful in dense graph areas because it \textit{``points directly at it without you being confused if it was this one or the one behind it.''}
However, 14 participants (P1-9, P11, P13, P15, P17, and P18) noted that the Embodied avatar could obstruct the graph. 
P11 stated that \textit{``when they were pointing, it happened that they were overlapping the point, so I couldn't see what was written until I had to go there,''} and P13 explained that \textit{``their body blocks the node ... and that's a distraction.''} In addition, five participants (P2, P9, P13, P15, and P16) described the avatar's movement as visually distracting or mental demanding, especially in hard tasks. P13 noted that \textit{``when the task is harder, there's more information going on,''} making distraction harder to tolerate.

\vspace{0.2em}\noindent\textbf{Depth-adaptive Cursor.} Participants generally described the Cursor as a simple, lightweight cue with less visual clutter than the avatar representations. Five participants (P1, P2, P9, P17, and P18) found it especially clear in simpler tasks. P1 commented that \textit{``just the cursor reduces the visual clutter, so it's easy to just follow a simple path,''} and P18 said \textit{``there was nothing else going on ... could just focus on the cursor.''} Two participants (P16 and P17) also found it helpful for tracking the node selection. P17 stated that, in the hard round, \textit{``the cursor made it easier to remember the pattern''} because \textit{``there wasn't anything in the way.''} 
However, 11 participants (P2, P4-8, P10, P12, P13, P15, and P16) found the cursor small and difficult to locate, especially across distant graph regions. P6 explained that \textit{``I had to kind of look around on the graph to try to find where the cursor was,''} and P7 stated that \textit{``sometimes it is hard to like follow where the cursor went.''} Three participants (P13, P14, and P18) also reported weaker collaborator presence. P14 described it as \textit{``feeling that I was just going through some document alone.''} with less \textit{``collaborative feeling.''} Similarly, P18 noted that \textit{``I didn't really think about intent,''} describing the cursor as \textit{``more just robotic''} with \textit{``less human connection.''}

%% file: sections/06_discussion.tex
\section{Discussion}

\subsection{Do we need avatar representations for asymmetric collaborative visual analytics?}

\vspace{0.2em}\noindent\textbf{Avatar representations enhance collaborator awareness and provide expressive embodied cues of analytical activity.} Our results confirm that avatar-based representations provide stronger awareness and co-presence than cursor cues across task complexities. Participants reported lower aloneness, greater notice of the collaborator, and higher collaborator awareness under avatar conditions. Moreover, avatars provided richer cues for interpreting the desktop collaborator's analytical activity within the workspace. Participants found the avatar's position, body orientation, and arm gestures useful for interpreting the collaborator's activity, with movement signaling transitions across graph regions while orientation and pointing clarified engagement with specific data elements. 

\vspace{0.2em}\noindent\textbf{Avatar embodiment provides clear attentional guidance, but may also introduce visual and cognitive overhead.} Although avatar representations enhanced collaborator awareness, participants preferred the Distanced avatar over the Cursor and Embodied avatar. During the interview, participants noted that the Distanced avatar's pointing gesture and indication ray made the collaborator's target easier to follow, while its stable position kept the avatar from interfering with the graph. In contrast, the Embodied avatar's movement and proximity sometimes obscured analytical content and distracted participants, suggesting that increasing embodiment does not necessarily improve the collaborative experience. 

\subsection{How do different representations affect recall task performance for VR user?}

\textbf{Avatar representations may better support recall of the central point of the analysis.} Across both rounds, avatar conditions generally yielded higher central-point scores than the Cursor, although the difference was significant only for the Embodied avatar in the easy round. Related findings in cognitive science provide a possible perspective. Cutica and Bucciarelli found that speaker's co-speech gestures improved listeners' recall of the conceptual content~\cite{cutica2008deep}, while Nirme \textit{et al.} showed that speech-synchronized gestures facilitated recall process~\cite{nirme2024early}, suggesting that visible and aligned bodily cues may support memory for overall meaning. In our setting, avatar cues may have similarly helped participants retain the central point of the analysis. In the hard round, however, greater task complexity and information load may have attenuated this benefit, resulting in only numerical differences between conditions.

\vspace{0.2em}\noindent\textbf{Representation type has limited effects on analytical detail recall and reasoning.} No significant differences were observed among representations for node sequence reconstruction, key-event recall, or outcome reasoning in either round. Unlike central-point recall, these measures required participants to retain specific events, their temporal order, and causal relationships. Avatar representations primarily conveyed the collaborator's attentional focus and transitions across the graph without encoding event content, sequence, or causal structure, potentially limiting support for detailed recall and reasoning. This interpretation is broadly compatible with Cutica and Bucciarelli's finding that co-speech gestures benefited memory for conceptual meaning more than surface-level details \cite{cutica2008deep}. Avatar cues may also have introduced competing effects. Embodied and directional cues helped participants follow the collaborator, while avatar movement and occlusion could introduce distraction and hinder detailed recall. These opposing effects may explain performance comparable to the Cursor condition despite clearer differences in collaborator awareness and subjective preference.

\vspace{0.2em}\noindent\textbf{Representations have limited overall effects on perceived workload, while the Distanced avatar shows a modest advantage.} The three representations show broadly comparable workload in both rounds. Nevertheless, Distanced avatar tended to receive slightly lower workload than Embodied avatar across several dimensions, with frustration being significantly lower in the hard round, suggesting that its stable peripheral presentation may impose less additional burden than repeated embodied relocation. At the same time, Distanced also received similar workload ratings to those of Cursor across most dimensions. Overall, Distanced appears to preserve the low workload of a lightweight Cursor while showing a modest, though not consistent, advantage over Embodied avatar.

\subsection{How does task complexity affect the effectiveness of different representations?}

\textbf{Distanced avatars show greater workload changes as task complexity increases.} In the hard round, the Distanced avatar resulted in significantly higher mental demand and effort than in the easy round. However, because workload measures were comparable across representations in the hard round, this increase may reflect the greater demands of tracking and integrating information as task complexity increased. This interpretation aligns with interview feedback, as participants continued to value the Distanced avatar's guidance in harder tasks despite the increased workload.

\vspace{0.2em}\noindent\textbf{Complex tasks increase physical and attentional cost for Embodied avatar.} In the hard task round, the Embodied avatar significantly increased participants' physical demand. This might be because a longer analytical sequence required users to adjust their viewpoint more frequently. Moreover, interview feedback reflects a similar issue: as the task became more complex, participants focused more on analytical content and became less tolerant of workspace-level avatar movement.

\vspace{0.2em}\noindent\textbf{Cursor cues become harder to trace in higher-complexity tasks.} The interview findings showed that participants generally perceived the Cursor as a lightweight cue that was easy to follow in simpler tasks, but more difficult to locate when the analytical sequence became longer and involved shifts across spatially distant graph areas. As task complexity increased, the Cursor's small visual form provided limited information about the direction of these transitions, making it harder for participants to recover the collaborator's current focus after losing track of it. 

\subsection{Design Implications}

\vspace{0.2em}\noindent\textbf{Match representation to the spatial scale of analytical transitions.} Representation choice should depend not only on overall task complexity, but also on how the collaborator's attention moves through the workspace. A Cursor may be sufficient for interactions within a stable local region, whereas a Distanced avatar can better communicate shifts across spatially separated targets. Embodied relocation is most relevant when movement between regions is itself part of the analytical meaning.

\vspace{0.2em}\noindent\textbf{Use Distanced avatars as persistent directional anchors.} A Distanced avatar provides a stable social and spatial reference while its orientation, arm movement, and pointing ray indicate changing targets. This continuity helps users recover the collaborator's focus after distant transitions without placing the avatar directly inside the visualization. Distanced avatars are therefore particularly suitable when analytical activity is distributed across the workspace but does not require the collaborator's body to occupy the data space.

\vspace{0.2em}\noindent\textbf{Decouple target indication from body relocation.} Desktop interactions should not be mapped directly onto full-body avatar movement. Local interactions within the same analytical region can be represented through orientation and gestures, while relocation can be reserved for meaningful shifts in graph region or analytical focus. This separation preserves the expressive value of embodied movement without making every interaction visually disruptive.

\vspace{0.2em}\noindent\textbf{Coordinate embodied movement with the visualization layout.} An Embodied avatar should adapt its placement and movement to the data layout and user's viewpoint, considering node density, important links, viewing direction, and potential occlusion. This allows the avatar to remain near the analytical context while minimizing interference with the information being examined.

\vspace{0.2em}\noindent\textbf{Separate awareness support from analytical memory support.} Avatar cues can strengthen collaborator awareness and communicate current analytical activity, but they do not necessarily preserve event order, causal relationships, or earlier reasoning steps. Systems should therefore complement collaborator representations with persistent traces, transition histories, or explicit links among referenced evidence. This distinction allows the representation to support social and attentional coordination while other interface elements maintain detailed analytical continuity.

%% file: sections/07_limitation.tex
\section{Limitations and Future Work}

\vspace{0.2em}\noindent\textbf{Broader collaborative visual analysis activities.} Our study provides a controlled foundation for examining how avatar-based representations convey collaborators' analytical activity in immersive visual analysis. We focused on graph browsing and analytical recall, with desktop activity limited to node selection and verbal narration. Future work can extend Desk2Avatar to tasks such as comparison, filtering, clustering, and hypothesis revision to examine these representations across broader visual analysis processes.

\vspace{0.2em}\noindent\textbf{Refined Embodied avatar design.} Desk2Avatar demonstrates how relocation, body orientation, near-target pointing, and gestures can represent desktop activity in VR. Future work can refine the Embodied avatar by adapting its movement, placement, gestures, and rendering to graph density, target visibility, and task complexity to reduce occlusion and distraction. Gaze cues, transparency, and partial-body representations could further preserve embodiment while improving coordination with the visualization.

\vspace{0.2em}\noindent\textbf{Bidirectional cross-device representation.} Our study focuses on unidirectional desktop-to-VR representation. Future work can extend Desk2Avatar to bidirectional representation by conveying VR users and their immersive activities on the desktop, supporting live cross-device collaboration and studies of coordination, cue interpretation, misunderstanding repair, and strategy adaptation.

%% file: sections/08_conclusion.tex
\section{Conclusion}

We explored how avatar-based representations support asymmetric desktop-VR collaborative visual analytics by conveying desktop collaborators and their analytical activity in VR. We designed and compared two avatar representations, Distanced and Embodied, with a depth-adaptive cursor baseline in a controlled graph-browsing user study with narrative recall across task complexities. Through quantitative and qualitative analyses, we found that avatar enhanced awareness and co-presence, while recall task performance remained broadly comparable across conditions. Participants generally preferred the Distanced avatar for its stable attentional guidance and low visual overhead, whereas the Embodied avatar demonstrated the expressive potential of spatially grounded body movement alongside design challenges in relocation, proximity, and occlusion. We further discuss implications for future avatar representation design, highlighting how avatar placement, movement, and occlusion should be balanced with collaborator presence, attentional guidance, and workspace readability.

%% file: ref.bib
@inproceedings{zhou2022depth,
  title={In-depth mouse: Integrating desktop mouse into virtual reality},
  author={Zhou, Qian and Fitzmaurice, George and Anderson, Fraser},
  booktitle={Proceedings of the 2022 CHI Conference on Human Factors in Computing Systems},
  pages={1--17},
  year={2022},
  url = {https://doi.org/10.1145/3491102.3501884},
  doi={10.1145/3491102.3501884}
}

@article{isenberg2011collaborative,
  title={Collaborative visualization: Definition, challenges, and research agenda},
  author={Isenberg, Petra and Elmqvist, Niklas and Scholtz, Jean and Cernea, Daniel and Ma, Kwan-Liu and Hagen, Hans},
  journal={Information Visualization},
  volume={10},
  number={4},
  pages={310--326},
  year={2011},
  publisher={SAGE Publications Sage UK: London, England},
  url = {https://doi.org/10.1177/1473871611412817},
  doi={10.1177/1473871611412817}
}

@inproceedings{badam2017supporting,
  title={Supporting Team-First Visual Analytics through Group Activity Representations.},
  author={Badam, Sriram Karthik and Zeng, Zehua and Wall, Emily and Endert, Alex and Elmqvist, Niklas},
  booktitle={Graphics Interface},
  pages={208--213},
  year={2017},
  url = {https://dl.acm.org/doi/10.5555/3141475.3141515},
  doi={10.5555/3141475.3141515}
}

@article{reski2022empirical,
  title={An empirical evaluation of asymmetric synchronous collaboration combining immersive and non-immersive interfaces within the context of immersive analytics},
  author={Reski, Nico and Alissandrakis, Aris and Kerren, Andreas},
  journal={Frontiers in Virtual Reality},
  volume={2},
  pages={743445},
  year={2022},
  publisher={Frontiers Media SA},
  url={https://doi.org/10.3389/frvir.2021.743445},
  doi={10.3389/frvir.2021.743445}
}

@inproceedings{tong2023towards,
  title={Towards an understanding of distributed asymmetric collaborative visualization on problem-solving},
  author={Tong, Wai and Xia, Meng and Wong, Kam Kwai and Bowman, Doug A and Pong, Ting-Chuen and Qu, Huamin and Yang, Yalong},
  booktitle={2023 IEEE Conference Virtual Reality and 3D User Interfaces (VR)},
  pages={387--397},
  year={2023},
  organization={IEEE},
  url={https://doi.org/10.1109/VR55154.2023.00054},
  doi={10.1109/VR55154.2023.00054}
}

@article{enriquez2024evaluating,
  title={Evaluating layout dimensionalities in pc+ vr asymmetric collaborative decision making},
  author={Enriquez, Daniel and Tong, Wai and North, Chris and Qu, Huamin and Yang, Yalong},
  journal={Proceedings of the ACM on Human-Computer Interaction},
  volume={8},
  number={ISS},
  pages={112--132},
  year={2024},
  publisher={ACM New York, NY, USA},
  url = {https://doi.org/10.1145/3698130},
  doi={10.1145/3698130}
}

@inproceedings{brehault2025systematic,
  title={A Systematic Literature Review to Characterize Asymmetric Interaction in Collaborative Systems},
  author={Br{\'e}hault, Victor and Dubois, Emmanuel and Prouzeau, Arnaud and Serrano, Marcos},
  booktitle={Proceedings of the 2025 CHI Conference on Human Factors in Computing Systems},
  pages={1--19},
  year={2025},
  url = {https://doi.org/10.1145/3706598.3713129},
  doi={10.1145/3706598.3713129}
}

@article{gutwin2002descriptive,
  title={A descriptive framework of workspace awareness for real-time groupware},
  author={Gutwin, Carl and Greenberg, Saul},
  journal={Computer Supported Cooperative Work (CSCW)},
  volume={11},
  number={3},
  pages={411--446},
  year={2002},
  publisher={Springer},
  url = {https://doi.org/10.1023/A:1021271517844},
  doi={10.1023/A:1021271517844}
}

@article{lee2020shared,
  title={Shared surfaces and spaces: Collaborative data visualisation in a co-located immersive environment},
  author={Lee, Benjamin and Hu, Xiaoyun and Cordeil, Maxime and Prouzeau, Arnaud and Jenny, Bernhard and Dwyer, Tim},
  journal={IEEE Transactions on Visualization and Computer Graphics},
  volume={27},
  number={2},
  pages={1171--1181},
  year={2020},
  publisher={IEEE},
  url={https://doi.org/10.1109/TVCG.2020.3030450},
  doi={10.1109/TVCG.2020.3030450}
}

@article{piumsomboon2019effects,
  title={The effects of sharing awareness cues in collaborative mixed reality},
  author={Piumsomboon, Thammathip and Dey, Arindam and Ens, Barrett and Lee, Gun and Billinghurst, Mark},
  journal={Frontiers in Robotics and AI},
  volume={6},
  pages={5},
  year={2019},
  publisher={Frontiers Media SA},
  url={https://doi.org/10.3389/frobt.2019.00005},
  doi={10.3389/frobt.2019.00005}
}

@inproceedings{saffo2023through,
  title={Through their eyes and in their shoes: Providing group awareness during collaboration across virtual reality and desktop platforms},
  author={Saffo, David and Batch, Andrea and Dunne, Cody and Elmqvist, Niklas},
  booktitle={Proceedings of the 2023 CHI Conference on Human Factors in Computing Systems},
  pages={1--15},
  year={2023},
  url = {https://doi.org/10.1145/3544548.3581093},
  doi={10.1145/3544548.3581093}
}

@article{welsford2021spectator,
  title={Spectator view: Enabling asymmetric interaction between hmd wearers and spectators with a large display},
  author={Welsford-Ackroyd, Finn and Chalmers, Andrew and Kuffner dos Anjos, Rafael and Medeiros, Daniel and Kim, Hyejin and Rhee, Taehyun},
  journal={Proceedings of the ACM on Human-Computer Interaction},
  volume={5},
  number={ISS},
  pages={1--17},
  year={2021},
  publisher={ACM New York, NY, USA},
  url = {https://doi.org/10.1145/3486951},
  doi={10.1145/3486951}
}

@inproceedings{seraji2022xvcollab,
  title={XVCollab: An immersive analytics tool for asymmetric collaboration across the virtuality spectrum},
  author={Seraji, Mohammad Rajabi and Stuerzlinger, Wolfgang},
  booktitle={2022 IEEE International Symposium on Mixed and Augmented Reality Adjunct (ISMAR-Adjunct)},
  pages={146--154},
  year={2022},
  organization={IEEE},
  url={https://doi.org/10.1109/ISMAR-Adjunct57072.2022.00035},
  doi={10.1109/ISMAR-Adjunct57072.2022.00035}
}

@inproceedings{gottsacker2025examining,
  title={Examining the effects of immersive and non-immersive presenter modalities on engagement and social interaction in co-located augmented presentations},
  author={Gottsacker, Matt and Chen, Mengyu and Saffo, David and Lu, Feiyu and Lee, Benjamin and MacIntyre, Blair},
  booktitle={Proceedings of the 2025 CHI Conference on Human Factors in Computing Systems},
  pages={1--19},
  year={2025},
  url = {https://doi.org/10.1145/3706598.3713346},
  doi={10.1145/3706598.3713346}
}

@article{sarasso2024shared,
  title={Shared attention in virtual immersive reality enhances electrophysiological correlates of implicit sensory learning},
  author={Sarasso, Pietro and Ronga, Irene and Piovesan, Francesca and Barbieri, Paolo and Del Fante, Elena and De Luca, Daniela and Bechis, Ludovico and Osello, Anna and Sacco, Katiuscia},
  journal={Scientific Reports},
  volume={14},
  number={1},
  pages={3767},
  year={2024},
  publisher={Nature Publishing Group UK London},
  url={https://doi.org/10.1038/s41598-024-53937-w},
  doi={10.1038/s41598-024-53937-w}
}

@inproceedings{balakrishnan2008visualizations,
  title={Do visualizations improve synchronous remote collaboration?},
  author={Balakrishnan, Aruna D and Fussell, Susan R and Kiesler, Sara},
  booktitle={Proceedings of the SIGCHI conference on human factors in computing systems},
  pages={1227--1236},
  year={2008},
  url = {https://doi.org/10.1145/1357054.1357246},
  doi={10.1145/1357054.1357246}
}

@article{mahyar2014supporting,
  title={Supporting communication and coordination in collaborative sensemaking},
  author={Mahyar, Narges and Tory, Melanie},
  journal={IEEE transactions on visualization and computer graphics},
  volume={20},
  number={12},
  pages={1633--1642},
  year={2014},
  publisher={IEEE},
  url={https://doi.org/10.1109/TVCG.2014.2346573},
  doi={10.1109/TVCG.2014.2346573}
}

@inproceedings{reski2020oh,
  title={“Oh, that’s where you are!”--Towards a Hybrid Asymmetric Collaborative Immersive Analytics System},
  author={Reski, Nico and Alissandrakis, Aris and Tyrkk{\"o}, Jukka and Kerren, Andreas},
  booktitle={Proceedings of the 11th Nordic Conference on Human-Computer Interaction: Shaping Experiences, Shaping Society},
  pages={1--12},
  year={2020},
  url = {https://doi.org/10.1145/3419249.3420102},
  doi={10.1145/3419249.3420102}
}

@article{burova2022distributed,
  title={Distributed asymmetric virtual reality in industrial context: enhancing the collaboration of geographically dispersed teams in the pipeline of maintenance method development and technical documentation creation},
  author={Burova, Alisa and Palma, Paulina Becerril and Truong, Phong and M{\"a}kel{\"a}, John and Heinonen, Hanna and Hakulinen, Jaakko and Ronkainen, Kimmo and Raisamo, Roope and Turunen, Markku and Siltanen, Sanni},
  journal={Applied Sciences},
  volume={12},
  number={8},
  pages={3728},
  year={2022},
  publisher={MDPI},
  url={https://doi.org/10.3390/app12083728},
  doi={10.3390/app12083728}
}

@inproceedings{prouzeau2018awareness,
  title={Awareness techniques to aid transitions between personal and shared workspaces in multi-display environments},
  author={Prouzeau, Arnaud and Bezerianos, Anastasia and Chapuis, Olivier},
  booktitle={Proceedings of the 2018 ACM International Conference on Interactive Surfaces and Spaces},
  pages={291--304},
  year={2018},
  url = {https://doi.org/10.1145/3279778.3279780},
  doi = {10.1145/3279778.3279780}
}

@article{coppens2024supporting,
  title={Supporting mixed-presence awareness across wall-sized displays using a tracking pipeline based on depth cameras},
  author={Coppens, Adrien and Hermen, Johannes and Schwartz, Lou and Moll, Christian and Maquil, Val{\'e}rie},
  journal={Proceedings of the ACM on Human-Computer Interaction},
  volume={8},
  number={EICS},
  pages={1--32},
  year={2024},
  publisher={ACM New York, NY, USA},
  url = {https://doi.org/10.1145/3664634},
  doi = {10.1145/3664634}
}

@inproceedings{bai2020user,
  title={A user study on mixed reality remote collaboration with eye gaze and hand gesture sharing},
  author={Bai, Huidong and Sasikumar, Prasanth and Yang, Jing and Billinghurst, Mark},
  booktitle={Proceedings of the 2020 CHI conference on human factors in computing systems},
  pages={1--13},
  year={2020},
  url = {https://doi.org/10.1145/3313831.3376550},
  doi = {10.1145/3313831.3376550}
}

@article{yang2020effects,
  title={The effects of spatial auditory and visual cues on mixed reality remote collaboration},
  author={Yang, Jing and Sasikumar, Prasanth and Bai, Huidong and Barde, Amit and S{\"o}r{\"o}s, G{\'a}bor and Billinghurst, Mark},
  journal={Journal on Multimodal User Interfaces},
  volume={14},
  number={4},
  pages={337--352},
  year={2020},
  publisher={Springer},
  url={https://doi.org/10.1007/s12193-020-00331-1},
  doi={10.1007/s12193-020-00331-1}
}

@article{rasmussen2022supporting,
  title={Supporting workspace awareness in remote assistance through a flexible multi-camera system and Augmented Reality awareness cues},
  author={Rasmussen, Troels and Feuchtner, Tiare and Huang, Weidong and Gr{\o}nb{\ae}k, Kaj},
  journal={Journal of Visual Communication and Image Representation},
  volume={89},
  pages={103655},
  year={2022},
  publisher={Elsevier},
  url = {https://doi.org/10.1016/j.jvcir.2022.103655},
  doi = {10.1016/j.jvcir.2022.103655}
}

@inproceedings{woodworth2022redirecting,
  title={Redirecting desktop interface input to animate cross-reality avatars},
  author={Woodworth, Jason W and Broussard, David and Borst, Christoph W},
  booktitle={2022 IEEE conference on virtual reality and 3D user interfaces (VR)},
  pages={843--851},
  year={2022},
  organization={IEEE},
  url={https://doi.org/10.1109/VR51125.2022.00106},
  doi={10.1109/VR51125.2022.00106}
}

@article{oh2018systematic,
  title={A systematic review of social presence: Definition, antecedents, and implications},
  author={Oh, Catherine S and Bailenson, Jeremy N and Welch, Gregory F},
  journal={Frontiers in Robotics and AI},
  volume={5},
  pages={114},
  year={2018},
  publisher={Frontiers Media SA},
  url={https://doi.org/10.3389/frobt.2018.00114},
  doi={10.3389/frobt.2018.00114}
}

@article{kyrlitsias2022social,
  title={Social interaction with agents and avatars in immersive virtual environments: A survey},
  author={Kyrlitsias, Christos and Michael-Grigoriou, Despina},
  journal={Frontiers in Virtual Reality},
  volume={2},
  pages={786665},
  year={2022},
  publisher={Frontiers Media SA},
  url={https://doi.org/10.3389/frvir.2021.786665},
  doi={10.3389/frvir.2021.786665}
}

@inproceedings{wei2022communication,
  title={Communication in immersive social virtual reality: A systematic review of 10 years’ studies},
  author={Wei, Xiaoying and Jin, Xiaofu and Fan, Mingming},
  booktitle={Proceedings of the tenth international symposium of Chinese CHI},
  pages={27--37},
  year={2022},
  url = {https://doi.org/10.1145/3565698.3565767},
  doi = {10.1145/3565698.3565767}
}

@inproceedings{lee2006task,
  title={Task taxonomy for graph visualization},
  author={Lee, Bongshin and Plaisant, Catherine and Parr, Cynthia Sims and Fekete, Jean-Daniel and Henry, Nathalie},
  booktitle={Proceedings of the 2006 AVI workshop on BEyond time and errors: novel evaluation methods for information visualization},
  pages={1--5},
  year={2006},
  url = {https://doi.org/10.1145/1168149.1168168},
  doi = {10.1145/1168149.1168168}
}

@article{bradley1958complete,
  title={Complete counterbalancing of immediate sequential effects in a Latin square design},
  author={Bradley, James V},
  journal={Journal of the American Statistical Association},
  volume={53},
  number={282},
  pages={525--528},
  year={1958},
  publisher={Taylor \& Francis},
  url={https://doi.org/10.1080/01621459.1958.10501456},
  doi={10.1080/01621459.1958.10501456}
}

@inproceedings{biocca2001networked,
  title={The networked minds measure of social presence: Pilot test of the factor structure and concurrent validity},
  author={Biocca, Frank and Harms, Chad and Gregg, Jenn},
  booktitle={4th annual international workshop on presence, Philadelphia, PA},
  pages={1--9},
  year={2001}
}

@inproceedings{hart2006nasa,
  title={NASA-task load index (NASA-TLX); 20 years later},
  author={Hart, Sandra G},
  booktitle={Proceedings of the human factors and ergonomics society annual meeting},
  volume={50},
  number={9},
  pages={904--908},
  year={2006},
  organization={Sage publications Sage CA: Los Angeles, CA},
  url={https://doi.org/10.1177/154193120605000909},
  doi={10.1177/154193120605000909}
}

@inproceedings{grinstein2007vast,
  title={VAST 2007 contest-blue iguanodon},
  author={Grinstein, Georges and Plaisant, Catherine and Laskowski, Sharon and O'Connell, Theresa and Scholtz, Jean and Whiting, Mark},
  booktitle={2007 IEEE Symposium on Visual Analytics Science and Technology},
  pages={231--232},
  year={2007},
  organization={IEEE},
  url={https://doi.org/10.1109/VAST.2007.4389032},
  doi={10.1109/VAST.2007.4389032}
}

@article{guo2023effects,
  title={The effects of visual complexity and task difficulty on the comprehensive cognitive efficiency of cluster separation tasks},
  author={Guo, Qi and Chen, Yan},
  journal={Behavioral Sciences},
  volume={13},
  number={10},
  pages={827},
  year={2023},
  publisher={MDPI},
  url={https://doi.org/10.3390/bs13100827},
  doi={10.3390/bs13100827}
}

@incollection{dwyer2018immersive,
  title={Immersive analytics: An introduction},
  author={Dwyer, Tim and Marriott, Kim and Isenberg, Tobias and Klein, Karsten and Riche, Nathalie and Schreiber, Falk and Stuerzlinger, Wolfgang and Thomas, Bruce H},
  booktitle={Immersive analytics},
  pages={1--23},
  year={2018},
  publisher={Springer},
  url={https://doi.org/10.1007/978-3-030-01388-2_1},
  doi={10.1007/978-3-030-01388-2_1}
}

@inproceedings{donalek2014immersive,
  title={Immersive and collaborative data visualization using virtual reality platforms},
  author={Donalek, Ciro and Djorgovski, S George and Cioc, Alex and Wang, Anwell and Zhang, Jerry and Lawler, Elizabeth and Yeh, Stacy and Mahabal, Ashish and Graham, Matthew and Drake, Andrew and others},
  booktitle={2014 IEEE International conference on big data (big data)},
  pages={609--614},
  year={2014},
  organization={IEEE},
  url={https://doi.org/10.1109/BigData.2014.7004282},
  doi={10.1109/BigData.2014.7004282}
}

@inproceedings{heidrich2021towards,
  title={Towards a Collaborative Experimental Environment for Graph Visualization Research in Virtual Reality.},
  author={Heidrich, David and Meinecke, Annika and Schreiber, Andreas and By{\v{s}}ka, J and J{\"a}nicke, S and Schmidt, J},
  booktitle={EuroVis (Posters)},
  pages={9--11},
  year={2021},
  url={https://doi.org/10.2312/evp.20211068},
  doi={10.2312/evp.20211068}
}

@inproceedings{zagermann2023challenges,
  title={Challenges and opportunities for collaborative immersive analytics with hybrid user interfaces},
  author={Zagermann, Johannes and Hubenschmid, Sebastian and Fink, Daniel Immanuel and Wieland, Jonathan and Reiterer, Harald and Feuchtner, Tiare},
  booktitle={2023 IEEE International Symposium on Mixed and Augmented Reality Adjunct (ISMAR-Adjunct)},
  pages={191--195},
  year={2023},
  organization={IEEE},
  url={https://doi.org/10.1109/ISMAR-Adjunct60411.2023.00044},
  doi={10.1109/ISMAR-Adjunct60411.2023.00044}
}

@book{hartson2012ux,
  title={The UX Book: Process and guidelines for ensuring a quality user experience},
  author={Hartson, Rex and Pyla, Pardha S},
  year={2012},
  publisher={Elsevier},
  url = {https://doi.org/10.1145/2559866.2559873},
  doi = {10.1145/2559866.2559873}
}

@book{laviola20173d,
  title={3D user interfaces: theory and practice},
  author={LaViola Jr, Joseph J and Kruijff, Ernst and McMahan, Ryan P and Bowman, Doug and Poupyrev, Ivan P},
  year={2017},
  publisher={Addison-Wesley Professional},
  url={https://dl.acm.org/doi/10.5555/993837},
  doi={10.5555/993837}
}

@article{mine1995virtual,
  title={Virtual environment interaction techniques},
  author={Mine, Mark R},
  journal={UNC Chapel Hill CS Dept},
  year={1995},
  url={https://dl.acm.org/doi/10.5555/897820},
  doi={10.5555/897820}
}

@inproceedings{levenshtein1966binary,
  title={Binary codes capable of correcting deletions, insertions, and reversals},
  author={Levenshtein, Vladimir I and others},
  booktitle={Soviet physics doklady},
  volume={10},
  number={8},
  pages={707--710},
  year={1966},
  organization={Soviet Union}
}

@article{holm1979simple,
  title={A simple sequentially rejective multiple test procedure},
  author={Holm, Sture},
  journal={Scandinavian journal of statistics},
  pages={65--70},
  year={1979},
  publisher={JSTOR}
}

@article{aickin1996adjusting,
  title={Adjusting for multiple testing when reporting research results: the Bonferroni vs Holm methods.},
  author={Aickin, Mikel and Gensler, Helen},
  journal={American journal of public health},
  volume={86},
  number={5},
  pages={726--728},
  year={1996},
  publisher={American Public Health Association},
  url={https://doi.org/10.2105/ajph.86.5.726},
  doi={10.2105/ajph.86.5.726}
}

@inproceedings{frohler2022survey,
  title={A Survey on Cross-Virtuality Analytics},
  author={Fr{\"o}hler, Bernhard and Anthes, Christoph and Pointecker, Fabian and Friedl, Judith and Schwajda, Daniel and Riegler, Andreas and Tripathi, Shailesh and Holzmann, Clemens and Brunner, Manuel and Jodlbauer, Herbert and others},
  booktitle={Computer graphics forum},
  volume={41},
  number={1},
  pages={465--494},
  year={2022},
  organization={Wiley Online Library},
  url={https://doi.org/10.1111/cgf.14447},
  doi={10.1111/cgf.14447}
}

@inproceedings{belcher2003using,
  title={Using augmented reality for visualizing complex graphs in three dimensions},
  author={Belcher, Daniel and Billinghurst, Mark and Hayes, SE and Stiles, Randy},
  booktitle={The Second IEEE and ACM International Symposium on Mixed and Augmented Reality, 2003. Proceedings.},
  pages={84--93},
  year={2003},
  organization={IEEE},
  url={https://doi.org/10.1109/ISMAR.2003.1240691},
  doi={10.1109/ISMAR.2003.1240691}
}

@inproceedings{ware1994viewing,
  title={Viewing a graph in a virtual reality display is three times as good as a 2D diagram},
  author={Ware, Colin and Franck, Glenn},
  booktitle={Proceedings of 1994 IEEE symposium on visual languages},
  pages={182--183},
  year={1994},
  organization={IEEE},
  url={https://doi.org/10.1109/VL.1994.363621},
  doi={10.1109/VL.1994.363621}
}

@article{ware1996evaluating,
  title={Evaluating stereo and motion cues for visualizing information nets in three dimensions},
  author={Ware, Colin and Franck, Glenn},
  journal={ACM Transactions on Graphics (TOG)},
  volume={15},
  number={2},
  pages={121--140},
  year={1996},
  publisher={ACM New York, NY, USA},
  url = {https://doi.org/10.1145/234972.234975},
  doi = {10.1145/234972.234975}
}

@inproceedings{gou2012socialnetsense,
  title={SocialNetSense: supporting sensemaking of social and structural features in networks with interactive visualization},
  author={Gou, Liang and Zhang, Xiaolong and Luo, Airong and Anderson, Patricia F},
  booktitle={2012 IEEE Conference on Visual Analytics Science and Technology (VAST)},
  pages={133--142},
  year={2012},
  organization={IEEE},
  url={https://doi.org/10.1109/VAST.2012.6400558},
  doi={10.1109/VAST.2012.6400558}
}

@inproceedings{zhao2025libra,
  title={Libra: An interaction model for data visualization},
  author={Zhao, Yue and Wang, Yunhai and Luo, Xu and Wang, Yanyan and Fekete, Jean-Daniel},
  booktitle={Proceedings of the 2025 CHI Conference on Human Factors in Computing Systems},
  pages={1--17},
  year={2025},
  url = {https://doi.org/10.1145/3706598.3713769},
  doi = {10.1145/3706598.3713769}
}

@article{cassell2001embodied,
  title={Embodied conversational agents: representation and intelligence in user interfaces},
  author={Cassell, Justine},
  journal={AI magazine},
  volume={22},
  number={4},
  pages={67--67},
  year={2001},
  url = {https://doi.org/10.1609/aimag.v22i4.1593},
  doi = {10.1609/aimag.v22i4.1593}
}

@inproceedings{casanueva2000effects,
  title={The effects of group collaboration on presence in a collaborative virtual environment},
  author={Casanueva, Juan and Blake, Edwin},
  booktitle={Virtual Environments 2000: Proceedings of the Eurographics Workshop in Amsterdam, The Netherlands, June 1--2, 2000},
  pages={85--94},
  year={2000},
  organization={Springer},
  url={https://dl.acm.org/doi/10.5555/2385930.2385942},
  doi={10.5555/2385930.2385942}
}

@incollection{schroeder2002social,
  title={Social interaction in virtual environments: Key issues, common themes, and a framework for research},
  author={Schroeder, Ralph},
  booktitle={The social life of avatars: Presence and interaction in shared virtual environments},
  pages={1--18},
  year={2002},
  publisher={Springer},
  url={https://doi.org/10.1007/978-1-4471-0277-9_1},
  doi={10.1007/978-1-4471-0277-9_1}
}

@article{yang2022towards,
  title={Towards immersive collaborative sensemaking},
  author={Yang, Ying and Dwyer, Tim and Wybrow, Michael and Lee, Benjamin and Cordeil, Maxime and Billinghurst, Mark and Thomas, Bruce H},
  journal={Proceedings of the ACM on Human-Computer Interaction},
  volume={6},
  number={ISS},
  pages={722--746},
  year={2022},
  publisher={ACM New York, NY, USA},
  url = {https://doi.org/10.1145/3567741},
  doi = {10.1145/3567741}
}

@article{chung2014visporter,
  title={VisPorter: facilitating information sharing for collaborative sensemaking on multiple displays},
  author={Chung, Haeyong and North, Chris and Self, Jessica Zeitz and Chu, Sharon and Quek, Francis},
  journal={Personal and Ubiquitous Computing},
  volume={18},
  number={5},
  pages={1169--1186},
  year={2014},
  publisher={Springer},
  url = {https://doi.org/10.1007/s00779-013-0727-2},
  doi = {10.1007/s00779-013-0727-2}
}

@inproceedings{badam2014polychrome,
  title={Polychrome: A cross-device framework for collaborative web visualization},
  author={Badam, Sriram Karthik and Elmqvist, Niklas},
  booktitle={Proceedings of the Ninth ACM International Conference on Interactive Tabletops and Surfaces},
  pages={109--118},
  year={2014},
  url = {https://doi.org/10.1145/2669485.2669518},
  doi = {10.1145/2669485.2669518}
}

@article{kerby2014simple,
  title={The simple difference formula: An approach to teaching nonparametric correlation},
  author={Kerby, Dave S},
  journal={Comprehensive Psychology},
  volume={3},
  pages={11--IT},
  year={2014},
  publisher={SAGE Publications Sage CA: Los Angeles, CA},
  url={https://doi.org/10.2466/11.IT.3.1},
  doi={10.2466/11.IT.3.1}
}

@article{yoon2021full,
  title={A full body avatar-based telepresence system for dissimilar spaces},
  author={Yoon, Leonard and Yang, Dongseok and Chung, Choongho and Lee, Sung-Hee},
  journal={arXiv preprint arXiv:2103.04380},
  year={2021}
}

@article{yang2024visual,
  title={Visual guidance for user placement in avatar-mediated telepresence between dissimilar spaces},
  author={Yang, Dongseok and Kang, Jiho and Kim, Taehei and Lee, Sung-Hee},
  journal={IEEE Transactions on Visualization and Computer Graphics},
  volume={30},
  number={12},
  pages={7558--7570},
  year={2024},
  publisher={IEEE},
  url={https://doi.org/10.1109/TVCG.2024.3354256},
  doi={10.1109/TVCG.2024.3354256}
}

@misc{wasserstein2016asa,
  title={The ASA statement on p-values: context, process, and purpose},
  author={Wasserstein, Ronald L and Lazar, Nicole A},
  journal={The American Statistician},
  volume={70},
  number={2},
  pages={129--133},
  year={2016},
  publisher={Taylor \& Francis},
  url={https://doi.org/10.1080/00031305.2016.1154108},
  doi={10.1080/00031305.2016.1154108}
}

@incollection{dragicevic2016fair,
  title={Fair statistical communication in HCI},
  author={Dragicevic, Pierre},
  booktitle={Modern statistical methods for HCI},
  pages={291--330},
  year={2016},
  publisher={Springer},
  url={https://doi.org/10.1007/978-3-319-26633-6_13},
  doi={10.1007/978-3-319-26633-6_13}
}

@article{schumm2013determining,
  title={Determining statistical significance (alpha) and reporting statistical trends: Controversies, issues, and facts},
  author={Schumm, Walter R and Pratt, Kariga K and Hartenstein, Jaimee L and Jenkins, Bertha A and Johnson, Gralon A},
  journal={Comprehensive Psychology},
  volume={2},
  pages={03--CP},
  year={2013},
  publisher={SAGE Publications Sage CA: Los Angeles, CA},
  url={https://doi.org/10.2466/03.CP.2.10},
  doi={10.2466/03.CP.2.10}
}

@inproceedings{hart2021manipulating,
  title={Manipulating avatars for enhanced communication in extended reality},
  author={Hart, Jonathon Derek and Piumsomboon, Thammathip and Lee, Gun A and Smith, Ross T and Billinghurst, Mark},
  booktitle={2021 IEEE International Conference on Intelligent Reality (ICIR)},
  pages={9--16},
  year={2021},
  organization={IEEE},
  url={https://doi.org/10.1109/ICIR51845.2021.00011},
  doi={10.1109/ICIR51845.2021.00011}
}

@article{cutica2008deep,
  title={The deep versus the shallow: Effects of co-speech gestures in learning from discourse},
  author={Cutica, Ilaria and Bucciarelli, Monica},
  journal={Cognitive science},
  volume={32},
  number={5},
  pages={921--935},
  year={2008},
  publisher={Wiley Online Library},
  url={https://doi.org/10.1080/03640210802222039},
  doi={10.1080/03640210802222039}
}

@article{nirme2024early,
  title={Early or synchronized gestures facilitate speech recall—A study based on motion capture data},
  author={Nirme, Jens and Gulz, Agneta and Haake, Magnus and Gullberg, Marianne},
  journal={Frontiers in Psychology},
  volume={15},
  pages={1345906},
  year={2024},
  publisher={Frontiers Media SA},
  url={https://doi.org/10.3389/fpsyg.2024.1345906},
  doi={10.3389/fpsyg.2024.1345906}
}
